# Super-transport of Excitons in Atomically Thin Organic Semiconductors at the 2D Quantum Limit

Ankur Sharma,[1] Linglong Zhang,[1] Jonathan O. Tollerud[2], Miheng Dong,[1] Yi Zhu,[1] Robert Halbich,[1] Tobias Vogl,[3] Kun Liang,[1,4], Hieu T. Nguyen,[1] Fan Wang,[5] Shilpa Sanwlani[2], Stuart K. Earl[2], Daniel Macdonald,[1] Ping Koy Lam,[3] Jeff A. Davis[2] and Yuerui Lu[1*]

[1]Research School of Engineering, College of Engineering and Computer Science, The Australian National University, Canberra, ACT, 2601, Australia

[2]Centre for Quantum and Optical Science, Swinburne University of Technology, Hawthorn, Victoria, 3122, Australia and ARC Centre of Excellence for Future Low-Energy Electronics Technology, Swinburne University of Technology, Hawthorn, Victoria, 3122, Australia

[3]Centre for Quantum Computation and Communication Technology, Department of Quantum Science, Research School of Physics and Engineering, The Australian National University, Acton ACT, 2601, Australia

[4]School of Mechatronical Engineering, Beijing Institute of Technology, Beijing 100081, China

[5]Institute for Biomedical Materials and Devices (IBMD), Faculty of Science, University of Technology Sydney, NSW 2007, Australia.

* To whom correspondence should be addressed: Yuerui Lu (yuerui.lu@anu.edu.au)

**Abstract:**

**Long-range and fast transport of coherent excitons is important for development of high-speed excitonic circuits and quantum computing applications. However, most of these coherent excitons have only been observed in some low-dimensional semiconductors when coupled with cavities, as there are large inhomogeneous broadening and dephasing effects on the exciton transport in their native states of the materials. Here, by confining coherent excitons at the 2D quantum limit, we firstly observed molecular aggregation enabled 'super-transport' of excitons in atomically thin two-dimensional (2D) organic semiconductors between coherent states, with a measured a high effective exciton diffusion coefficient of ~346.9 cm$^2$/sec at room temperature. This value is one to several orders of magnitude higher than the reported values from other organic molecular aggregates and low-dimensional inorganic materials. Without coupling to any optical cavities, the monolayer pentacene sample, a very clean 2D quantum system (~1.2 nm**

**thick) with high crystallinity (J-type aggregation) and minimal interfacial states, showed superradiant emissions from the Frenkel excitons, which was experimentally confirmed by the temperature-dependent photoluminescence (PL) emission, highly enhanced radiative decay rate, significantly narrowed PL peak width and strongly directional in-plane emission. The coherence in monolayer pentacene samples was observed to be delocalized over ~135 molecules, which is significantly larger than the values (a few molecules) observed from other organic thin films. In addition, the super-transport of excitons in monolayer pentacene samples showed highly anisotropic behaviour. Our results pave the way for the development of future high-speed excitonic circuits, fast OLEDs, and other opto-electronic devices.**

There has been an increasing interest in recent times to harness the long-range and fast transport of coherent excitons (electron-hole pair quasi-particles) in solid state inorganic semiconductors and molecular systems with confined geometries for enhancing light-matter interactions. [1, 2, 3] Coherence is critical to engineer quantum electrodynamics (QED)[4] in low-dimensional systems, such as quantum wells[5], two-dimensional (2D) materials[6] and quantum dots[7], leading to various applications in realizing quantum memories,[8] single-photon sources,[9] laser cooling,[10] narrow linewidth lasers,[11] high efficiency solar cells,[12] and stable polaritons to achieve Bose Einstein condensates.[5] Long-range and fast migration of coherent excitons also have tremendous applications in schemes of high-speed excitonic circuits, quantum computing,[16-18] and high quantum yield light emitting diodes.[19]

A spontaneous coherent emission from a system of several non-interacting dipole active atoms (excitons) was defined as superradiance (SR) by Dicke.[20] In this phenomenon interactions between transition dipoles of individual molecules allow coherent delocalization across multiple sites. This leads to a net enhancement in the optical transitional dipole moment (TDM) value and sharp enhancement of the excitonic radiative decay rate of an ensemble of $N_c$ independent emitters as compared to the radiation decay rate of a single emitter.[2,7] The same principle of coherent delocalized superradiant emission gives rise to an analogous phenomenon called cooperative energy transfer or supertransfer (ST).[21] The resulting enhanced oscillator strength from delocalization over large molecular assemblies can lead to large scale exciton transport. In a ST process, the molecular assemblies (consisting of $N_c$ molecules) with comparable net TDMs can play the role of acceptors like individual molecules in Förster resonance energy transfer (FRET) mechanism.[22] Thus, the excitation (exciton) can transfer to much longer distances in a delocalized molecular assembly before annihilation, resulting in large values of effective exciton diffusion coefficients.

Recently, coherent exciton transport with highly enhanced effective exciton diffusion coefficients (3-70 cm$^2$/s)[23-25] was observed in quasi-one-dimensional (quasi-1D) molecular assemblies (cylindrical bundles,[23] nano-tubes[22] and wires[24,25]), which was significantly higher than the migration speeds of incoherent excitons reported in other materials, including conventional organic thin films and their heterostructures (0.001- 3 cm$^2$/s)[26,27], III–V semiconductor quantum wells (0.1-10 cm$^2$/s)[28], atomically thin transition metal dichalcogenides (TMDs) and phosphorene (0.01-14.5 cm$^2$/s)[29], *etc*. (Table S1 in Supporting Information). However, the migration speeds of the coherent excitons in these quasi-1D systems[23-25] were still hindered by the low overall exciton oscillation strengths (low coherence length with small $N_c$ values), which were mainly limited by the low quantum confinement of excitons[30] (diameter size > 12 nm) and potential disorders and interfacial sites in the systems.[31] Also, those quasi-1D systems have very small cross-section area for light-matter interactions, limiting their applications for future excitonic devices. [3,13-15] As predicted by theory, exciton oscillation strength, a key parameter for coherent excitons, is very sensitive to the quantum confinement of the system.[30] Therefore, confining the coherent excitons at the 2D quantum limit will be a very promising way to realize the long-range and fast transport of excitons;

moreover, the 2D structure can also provide a large cross section area for light-matter interactions, enabling tremendous applications in future excitonic devices.

In this work, from atomically thin organic molecular crystals without any optical cavities, we observed a long-range and fast migration of Frenkel (FR) excitons between coherent states at the 2D quantum limit. A very high effective exciton diffusion coefficient of ~346.9 cm$^2$/sec at room temperature was reported, which is one order of magnitude higher than the previously reported values (Table S1 in Supporting Information) from other materials.[23,25,32] The sharp, strong emission from J-type (monolayer, short as 1L) aggregation in pentacene in contrast to the emissions from H-type (wetting layer, short as WL) aggregates is confirmed to be the superradiant emission from the FR excitons. The coherence in 1L pentacene samples was determined to be delocalized over ~135 molecules, which is more than one order of magnitude larger than the values (a few molecules) observed from other organic thin films.[33] Our simulation results from quantum calculations attribute this to the constructive dipole coupling in J-type aggregation in 1L pentacene to form an enhanced net optical dipole moment, which supports our experimental observations. The super-transport of excitons observed in the macroscopic scale regime will enable the realization of strongly enhanced light-matter interactions at the quantum limit using such 2D organic materials, which has key applications in developing high-speed quantum computing devices, fast response time OLEDs, excitonic transistors and other opto-electronic devices.[8,15,18]

**Results**

**Sample growth and characterization**

Single-crystal pentacene (Figure 1) was epitaxially grown layer by layer on a hexagonal boron nitride (h-BN) surface with atomic smoothness and well-defined crystal facets. The first pentacene layer on h-BN was named as the wetting layer (WL; ~0.6 nm thick) and the next layer of pentacene grown on WL was designated as monolayer (1L; ~1.2 nm thick)[34] (Figure 1a-c). The layer-dependent arrangement of pentacene molecules over h-BN and their electrical transport properties were described in recent report[34]. Raman spectroscopy was used to confirm the presence of pentacene on h-BN (Figure S1). PL spectroscopy with a continuous-wave 532 nm excitation laser was used to characterize the excitonic emissions from those atomically thin pentacene layers. The WL region showed much stronger PL emissions than the 1L regions at room temperature (Figure 1d and Figure S2 and associated text). Compared with inorganic TMD 2D semiconductors, the WL showed a much broader PL spectrum at room temperature, which is associated with various band energy levels formed in pentacene due to vibronic couplings between FR and charge-transfer (CT) excitons (Supporting Information note 1 and Figure S3)[35,36]. On the other hand, 1L region showed a single peak centred around 680 nm that is predominantly FR excitonic emission, which will be explained later. This region is a heterostructure consisting of 1L on top of WL. However, in the PL spectrum from the 1L region, we did not see the higher-energy emission (~550-650 nm) shown in WL sample (Figure 1d), because the photo-excited charges in WL underneath could be quickly transferred to 1L before radiative emission occurred (Figure S2). Therefore, the PL spectrum measured from 1L region is coming from 1L pentacene sample and reflects its intrinsic nature.

**Superradiance and excitonic characteristics**

To explore the nature of excitonic transitions, we conducted temperature-dependent PL measurements on the WL and 1L pentacene samples down to 77 K (Figure 2). With decreasing temperature, the PL intensity from WL dropped significantly and the full-width half maximum (FWHM) broadened (Figure 2a); in contrast, 1L pentacene showed the sharp rise of the PL intensity and a significantly reduced FWHM with decreasing temperature (Figures 2b and

Figure S4). The opposing temperature dependences were caused by the completely different molecular packings in WL and 1L pentacene, which can be fully understood from the theory of H and J molecular aggregation and their effects on optical emissions[35,37] (Supporting Information note 2; Figures S3). Theoretical values of coupling coefficients (*J*) and net TDM in each layer were extracted based on its molecular aggregation pattern and coupling (Figure S3a). WL follows a parallel packing of TDMs, and 1L follows a head-to-tail packing of TDMs because of the herringbone packing of pentacene molecules in the unit cell (Figure 1c, and S3)[38]. *J* values for 1L and WL pentacene are negative and positive, respectively, which suggest J- and H-type aggregations for 1L and WL, respectively.[38] The total TDM and oscillator strength for the lowest excited states ($S_1$) in WL are both zero, which lead to non-luminescent $S_1$ states at a temperature of absolute zero. In contrast, the total TDM for $S_1$ in 1L has a large value (with direction along the *b* axis of its unit cell), and its corresponding oscillator strength is high (Figure S3a). This resulted in luminescent $S_1$ states at absolute zero. The large magnitude of net TDM corresponds the phase-locking and constructive coupling of TDMs in 1L pentacene cooperatively resulting in the sharp rise in PL intensity and reduction in FWHM, which are characteristics of superradiant emission[33,35,39].

The phenomenon of superradiance is the combined process of coherent emissions from a cluster of $N_c$ molecular aggregates/chromophores whose oscillator strength is enhanced compared to a single emitter by a factor of $N_c$ owing to the large TDM associated with the molecular assembly[33,37]. To find which coherent size gives best agreement with experimental temperature-dependence of PL intensity, we used Huang-Rhys factor and the spectral strength of 0-0 and 0-1 transitions in the 1L PL spectrum to accurately determine the coherence number and hence the effect of vibrations and phononic interactions on the coherence length.[40] The $N_c$ value can also be estimated by the sharp reduction in FWHM and spectral strengths of the PL emission as the temperature decreases.[39] Based on this technique and our measurements in

Figure 2c, S4 and S5, we estimated this value to be over 135 molecules. (See Supporting information Note 3). The extracted $N_c$ value from our 1L pentacene sample is more than one order of magnitude larger than that observed from tetracene thin film ($N_c$ ~ a few molecules).[33] The $N_c$ value signifies the extent of FR exciton delocalization over 135 consecutive dipoles in 1L pentacene. The enhanced oscillator strength is distributed over these consecutive dipoles and results in a large net TDM as shown in Figure S3.

Excitonic superradiance also leads to shortening of excitation radiative decay lifetime for the emission[2,4,7,11,33,39] by a factor of $N_c$. To confirm this coherent superradiance from 1L, we measured the temperature-dependent lifetimes of two excitonic emissions from WL and 1L pentacene samples using time resolved photoluminescence (TRPL) measurements with a resolution of 2.1 ps after deconvolution (Figure S6). The lifetime observed from 1L pentacene sample significantly dropped from ~12.7 ± 0.8 ps at room temperature down to ~ 4.1 ± 0.7 ps at 77 K (Figure 2c). To further confirm the measurement accuracy of the lifetime, we performed time and spectrally resolved differential transient reflectance (TR) measurements on WL and 1L pentacene samples at 77 K using a pump-probe technique with a time resolution of 200 fs. 1L pentacene sample showed a clear dip in the differential TR mapping centred at 677 nm; in contrast, the dip from WL sample was not clear (Figure 2d-e). The relaxation of this feature in 1L sample follows a fast decay (lifetime of ~4.2 ± 0.53 ps) (Figure 2f). The fitting of the decay curve was done using a single exponential decay and a 95% confidence interval, giving an estimated lifetime range of 3.6-4.7 ps (Figure S7). Further details about the measurements and the fitting are explained in supporting information Figure S7 and associated text. The consistency between the two complementary techniques confirms the accuracy of our lifetime values measured from our TRPL system. With the decrease of temperature, the significant reduction of emission lifetime was consistent with the observed sharp rise of the PL intensity in 1L pentacene sample (Figure 2b), confirming the fast and radiative nature the decay as expected for superradiance.

Further, the radiative lifetime of singlet FR excitonic emissions in bulk pentacene thin films (which are the non-coherent monomer states) was experimentally established to be ~1200 ps.[41] We also confirmed the lifetime from our bulk thin film pentacene grown on the same h-BN substrate, which was measured to be 1241.1 ps at room temperature (Figure S8 and associated text). The significantly reduced lifetime (~12.7 ps) in our 1L pentacene sample as compared to the value (~1200 ps) from the non-coherent states in thin film pentacene, coupled with the consistent temperature-dependent PL emission and lifetime, further substantiates superradiant emission from 1L pentacene. The value of $N_c$, as extracted from this sharp reduction in lifetime is ~100 at room temperature, which is again in close agreement with the aforementioned extracted values based the highly-enhanced PL intensity and reduced FWHM with the decrease of temperature. In contrast, the lifetimes from WL pentacene sample were measured to be 2.89 ns using TRPL at room temperature and 2.64 ns at 77 K (Figure 2c), much larger than the values from 1L pentacene samples, confirming the non-coherent nature of the CT excitons in WL samples.

We further performed PL measurements with angle-resolved excitation and emission polarizations (Figure 2g and S9) to confirm the anisotropic nature of superradiant FR excitonic emissions from 1L. PL intensity from 1L strongly depended on the emission polarization angle $\theta$ and showed a period of 180 degrees; on the contrary, the emission from WL was isotropic (Figure 2g). The strongly directional total TDM in 1L pentacene theoretically predicted (Figure S3a) is expected to lead to an in-plane anisotropic emission of those FR excitons. Henceforth, the zero-degree reference in the polar plots (Figure 2f) and subsequent figures can be identified as the *b* axis of the unit cell in 1L pentacene (Figure 1c and S3a). The directional emission in 1L pentacene sample was in-dependent of the excitation polarization angle of incident laser

(Figure S9). Similar anisotropy for PL emission from 1L pentacene was observed from 1L samples at room temperature (Figure S10). Normally, the dichroic ratio (*DR*) is a term used to characterize the in-plane anisotropy of optical transitions[42]. Here, it is defined as $DR = I_\parallel /I_\perp$, where $I_\parallel$ and $I_\perp$ are the PL peak intensities measured at polarization angles of 0° and 90°, respectively. The measured *DR* values of PL emission from 1L for emission (Figure 2g-blue curve) polarizations at 77 K were high up at 10.3, which indicates the strongly anisotropic nature of PL emission from 1L. At room temperature the *DR* value observed was close to 5.9 from 1L samples. In sharp contrast, the measured *DR* values from PL emission in WL pentacene were isotropic both at room temperature and 77 K (Figure 2g-red curve), which suggests that these peaks are dominated by the contribution of CT excitons.[43] Supporting Information Note 4 showed a further detailed discussion on the different nature of excitonic states in WL and 1L.

**Long-range and superfast exciton transport**

The same principle of coherent delocalized superradiant emission gives rise to an analogous phenomenon called cooperative energy transfer or supertransfer.[21] The resulting enhanced oscillator strength from delocalization over large molecular assemblies can lead to large scale exciton transport. The near-field direct imaging technique with a charge-coupled device (CCD) camera[24,25,44] was used to detect the diffusion length of bright excitons from WL and 1L pentacene samples (Figure 3). Figure 3a (3b) shows the measured contour plots of PL intensity as a function of emission wavelength and space of exciton diffusion from 1L (WL) pentacene sample at 77 K, respectively. The spatial distribution of PL intensity can be extracted at the specific energies (shown by dotted lines in Figure 3a and 3b). Figure 3c shows the extent of spatial diffusion in a steady 1D diffusion model. A gaussian fitting routine[24,44] was used to quantitatively compare the reflected excitation beam to the emission profile emerging from pentacene. Both emission and laser profiles were fitted using a bivariate normal distribution and their FWHMs were extracted. The spatial extent of exciton diffusion (diffusion length $L_D$) was extracted by taking the difference between the standard deviations of the emission image and of the passive laser spot after fitting the raw data. A gaussian fitting model (See methods) was used as the excitation from laser profile was predominantly gaussian and the diffusion model was not pair mediated or defect assisted in our high crystalline sample.[45,46] The exciton diffusion lengths ($L_D$) in 1L and WL pentacene samples were measured to be 0.55 ± 0.06 μm and 1.36 ± 0.16 μm, respectively (Figure 3d). All the measurements were taken at low excitation power of around 6 μW. We also did measurements with different power ranging from 6-215 μW and similar diffusion lengths and lifetimes were observed, which shows the minimized and ignorable effect of exciton-exciton annihilation in our measurements (Figure S11).[22,47] Using the measured lifetimes of excitonic emissions ($\tau$), the effective exciton diffusion coefficient ($D_{eff}$) could be extracted using the equation[25,47] $L_D = \sqrt{2D_{eff}\tau}$. Using the lifetimes (Figure 2e), the extracted diffusion coefficients from 1L and WL pentacene samples at 77 K were 354.5 ± 50.1 cm$^2$/sec and 3.5 ± 0.2 cm$^2$/sec, respectively (Figure 3d). The $D_{eff}$ value obtained from 1L is almost one order of magnitude higher than the reported values from other organic/inorganic systems with coherent and delocalized excitonic emissions (3-70 cm$^2$/sec, Table 1 in Supporting Information).[23-25] At the same time this value is around 3 orders of magnitude higher than value of diffusion constant reported for non-coherent singlet excitons in bulk thin film pentacene by Marciniak *et al.*[41] The exciton diffusion coefficients in WL and 1L at 77 K were further corroborated by another alternative method, spatial-temporal mapping[48] which is a very robust method to accurately determine exciton diffusion coefficient in similar materials.[48] Figure 4a shows the experimental setup used for spatial-temporal mapping and the obtained data plots stitched together are shown in Figure 4b, c. (See methods)

The extracted diffusion coefficient from the spatial-temporal mapping was obtained by extracting the gaussian dispersion of the emission profile over time is also consistent with our reported values obtained from time-resolved PL and diffusion mapping in space technique, highlighted in Figure 3. The obtained diffusion coefficient from 1L and WL is 306.8 ± 14.1 cm$^2$/sec and 3.3 ± 1.1 cm$^2$/sec viz. as obtained from the slope in Figure 4 d, e. It is important to remark here that the initial diffusion value is non-zero from WL due to cooling of hot-CT excitons in the <4 ps regime.[49] As discussed earlier in this report the WL is predominantly CT exciton dominated. For CT excitons in the initial few hundred femtoseconds to few ps regimes, the CT excitons undergo cooling-or loose excessive energy due to the creation of lot intermediate energy states between S$_1$ (natural excited exciton energy state) and CT states and they are all electronic states-- for excitons to quickly jump to in a few ps.[27,49] Further due to exciton residing on two different molecules, the extent of movement is also enhanced. All of this happens in a few hundred femtoseconds to few ps, hence the non-zero value in WL can be attributed to hot CT excitons states cooling rapidly to intermediate energy levels being created due to *H* type aggregation and donor/acceptor mediated hopping of excitons. Once that initial cooling happens the excitons then stabilize and then diffuse normally. Since, this non-zero transport occurs within a few ps regime, it does not affect the accuracy of the measurement of our exciton diffusion coefficient in WL, where the diffusion stretched to a few hundred ps-ns region. In 1L samples, which is predominantly FR exciton dominated, we did not observe such non-zero transport at *t=0*, as the hot CT excitons are practically non-existent there. The results further confirm the effectiveness of our technique to measure the precise exciton diffusion with a high resolution and limited role of triplet fission in our samples.

We attribute this high value of exciton diffusion coefficient in 1L to as the *supertransfer* phenomenon. The reason we were able to achieve these large effective exciton diffusion coefficients is three-fold. Firstly, in our 1L pentacene samples, the strong intermolecular coupling (J-type aggregation) results in a strong net exciton oscillator strength, and thus the delocalized and superradiant excitonic emissions. Within each delocalized segment, the excitation energy propagates ballistically[21], compared to the diffusive hopping described FRET. Delocalized excitons in principle accelerate energy transfer as compared to incoherent molecular hopping transfer because delocalization can define an effective hopping length that can be much larger than nearest inter-molecular spacings (Supporting Information note 5).[50-52] Secondly, because of the high-quality epitaxial 2D growth, our pentacene samples have highly-ordered crystallinity and minimized defects/disorders, resulting in long delocalization lengths and large coherence lengths ($N_c$ ~ 135 as shown in Figure S5, extracted from temperature dependent PL line strengths measurements), which significantly benefited the transport speed of coherent excitons. The exciton delocalization length is defined by competition between intermolecular coupling and disorders[22], which can be clearly seen from the comparison between PL spectra from high-quality and low-quality samples. In some 1L samples with low-quality growths, the FR exciton peak at ~680 nm was weak; instead, the high energy CT peaks would show up in the PL spectra (Figure S12; Supporting Information note 6). Compared with the high-quality 1L pentacene samples as shown in Figure 2, those low-quality 1L samples showed broader peak and lower PL peak intensity at 680 nm, and longer radiative lifetimes of the FR excitons (Figure S13), which is caused by the reduced coherence length in the low-quality 1L samples. In experiment, we also found that the disorders and interfacial states would increase as sample thickness increases, which leads to the de-coherence and reduced exciton delocalization length in thick samples, like 2L and bulk samples (both have J-aggregation). The 2L pentacene sample also showed the sharp emission peak at 680 nm, but the peak has slightly wider width and much lower PL intensity as compared to 1L (Figure S14). Whereas, bulk pentacene showed a broad PL emission and the 680 nm emission peak is almost negligible (Figure S8), due to the significantly reduced coherence length caused by the disorders and interfacial states in the sample, similar to other bulk organic thin films.[33] Of course, our 1L samples are still not perfect. There is still lots of room to improve the sample quality, which might give us higher coherence length and longer exciton diffusion length. Thirdly, the quantum confinement of the system will be beneficial to the speed of exciton transport, since the confinement can increase the exciton oscillation strength[30] (Supporting information note 7). Our 1L pentacene samples (~1.2 nm thick) have the intrinsic high confinement at the 2D quantum limit. Moreover, the excitons in 1L are highly anisotropic and can be aligned in a quasi-1D space along the *b* axis, which further increases the confinement of exciton transport in 1L pentacene. In summary, structural uniformity with defect free interface, highly confined excitons and excellent intermolecular coupling (large oscillator strengths) results in the supertransport of excitons in 1L pentacene. On the other hand, the diffusion coefficient (3.5 cm$^2$/sec; Figure 3d) of the CT excitons in WL sample is also much larger than that of other similar CT excitons reported recently[27]. Even though there is no coherence observed in WL, the strong spatial confinement of the excitons (~0.6 nm thick) and the almost defect-free interfaces in WL samples, lead to the long and fast propagation of long-lived CT excitons.[24] On the other hand the high diffusion length which is obtained in WL PEN samples is critical for applications in making high efficiency solar cells and charge separation.[53] In addition, the waveguiding effects can be excluded, as the thickness of the thickness of 1L and WL pentacene samples were <1.5 nm and the h-BN layer underneath was only around 4 nm (measured by AFM), thus making it insufficient for any light trapping and waveguiding, similar to previous reports.[25]

It is important to remark here that the observed diffusion was from bright singlet excitons in our 1L and WL pentacene samples and the long-lived triplet excitons *via* the singlet fission in pentacene would not affect the diffusion measurements (Figure 3). The reasons are two folds. Firstly, the triplet states are non-radiative dark states[54,55] and cannot be detected by the confocal photoluminescence setup we employed for our measurements.[32] Secondly, the regeneration of emissive singlets *via* subsequent triplet fusion is not allowed in pentacene and singlet fission in pentacene is expected to be exothermic and unidirectional[56,57] because the relaxed triplet in pentacene has significantly less than half the energy of the singlet. This is in contrast to the situation in tetracene, where the near-degenerate singlet and triplet-pair energy levels permit both the singlet fission and the triplet fusion to occur.

It is also important to remark here that the observed supertransport of excitons in 1L pentacene should be related to the unique properties of single fission in pentacene. Firstly, singlet fission in pentacene, a fast non-radiative decay, occurs within a time scale of ~80 fs.[57] On the picosecond time scale, the singlet fission in pentacene thin films has a very small total yield of approximately 2%.[41] In our 1L pentacene sample, both our pump-probe (Figure 2d) and TRPL (Figure 2c inset) measurements showed a mono-exponential radiative decay of ~4 ps, which indicates the minimal contribution of non-radiative singlet fission mechanisms in the 1-10 ps regime, consistent with previous report.[41] Secondly, our 1L pentacene sample at 77 K showed much stronger PL intensity than our 2L and bulk pentacene samples (Figure 2b, Figure S14 and S8), which suggests a highly reduced singlet fission rate in our 1L sample. This could be attributed to the minimized disorders in our 1L samples and the impact of the molecule packing on singlet fission in organic molecules.[58-61] Of course, it would be a very interesting future topic to further explore the aggregation and layer-dependent singlet fission in 2D pentacene samples.

**Superfast and angle-dependent transport of coherent excitons at room temperature**

Semiconductors with the long-range and fast transport of excitons at room temperature are critical for future high-speed excitonic circuits and quantum computing devices [16-18] that can operate at room temperature. Here, we also observed the superfast and angle-dependent transport of excitons from 1L pentacene samples at room temperature (Figure 5). In our measurement system, the polarization angle of incident laser was fixed to be parallel to the diffusion mapping direction. The exciton diffusion lengths and effective diffusion coefficients along two axes (*b* and *a*, labelled as 0° and 90°, respectively) of pentacene unit cell were measured, by rotating the sample with respect to the polarization of incident laser. The measured exciton diffusion length from 1L pentacene sample showed a maximum value of 0.93 ± 0.09 μm along the *b* axis (0°), and a minimum value of 0.37 ± 0.02 μm along *a* axis (90°) (Figure 5a and 5c). In contrast, the measured exciton diffusion length from WL sample was not angle-dependent, showing a constant value of ~1.31 ± 0.2 μm (Figure 5b and 5d). The exciton emission lifetime measured from 1L pentacene sample changed from ~12.7 ps at polarization angle of 0° to ~5.6 ps at polarization angle of 90° (Figure 4e). Whereas, the measured lifetime from WL pentacene sample remained largely unchanged (~2.8 ns) with changing the incident polarization angle (Figure 5e). The effective exciton diffusion coefficients were thus extracted and shown in Figure 5f. 1L pentacene sample showed a clear anisotropic exciton transport, with a very high effective diffusion coefficient of 346.9 ± 24.1 $cm^2$/sec along *b* axis and 95.3 ± 10.2 $cm^2$/sec along *a* axis (Figure 5f). Whereas, the WL shows a consistent effective diffusion coefficient of ~2.4 ± 0.1 $cm^2$/sec along different axes of the sample, due to its isotropic excitonic emission behaviour.

**Conclusion**

In conclusion, from high-quality atomically thin organic 2D semiconductors, we observed the supertransport of excitons between coherent excitonic states with a large effective diffusion coefficient of 346.9 cm$^2$/sec at room temperature, which is around one to several orders of magnitude higher than the reported values from other organic quasi-1D and thin film molecular aggregates[22-24]. We successfully identified molecular aggregation-sensitive PL emissions in atomically thin single-crystal organic semiconductors with different types of aggregation (H- and J-type) at the 2D quantum limit. The 1L pentacene samples with J-type aggregation (FR exciton dominated) showed superradiant emission, which was experimentally confirmed by the temperature-dependent PL emission, highly enhanced radiative decay rate, significantly narrowed PL peak width and strongly directional in-plane emission. This can be attributed to the constructive dipole coupling in J-type aggregation in 1L pentacene to form an enhanced net optical dipole moment. The excitons in 1L pentacene samples were extracted to be delocalized over ~135 molecules. In addition, the super-transport of excitons in monolayer pentacene samples showed highly anisotropic behaviour. On the other hand, the WL pentacene samples with H-type aggregation (CT exciton dominated) showed a fast migration of isotropic CT excitons with a measured diffusion coefficient of ~2.4 cm$^2$/sec at room temperature, a few times larger than that of other similar CT excitons reported[27]. Even though there is no coherence observed in WL, the strong spatial confinement of the excitons (~0.6 nm thick) and the almost defect-free interfaces in WL samples, lead to the long and fast propagation of long-lived CT excitons at the 2D quantum limit.[24] Our results provided the important experimental demonstration of the long-range and superfast transport of excitons between coherent states at the 2D quantum limit, paving the way for promising applications for future high-speed excitonic circuits, quantum computing devices, fast OLEDs, and other opto-electronic devices.

**Methods**

*Material growth*: h-BN flakes were mechanically exfoliated onto a thermally grown 285 nm thick $SiO_2$ layer deposited over silicon substrate. Before physical vapour deposition (PVD), optical microscope was used to characterize the topological information. The pentacene (purchased from Chem Supply: P0030-1G) was then deposited over the h-BN flakes, kept centred in a vacuum tube in the furnace. h-BN sheet on $SiO_2$/Si substrate was placed around 15 cm downstream and a molecular pump was used to evacuate the quartz tube to ~$10^{-4}$ mbar. The furnace was heated up to 130~150°C for 15~30 mins to grow pentacene. Then, the whole system was naturally cooled down to room temperature under vacuum. The number of layers were optimized by controlling source temperature, growth time and substrate position. All the samples were characterized, and layer thickness were identified using the standard AFM measurements, which were collected in ambient atmosphere at room temperature with a Bruker Multi-Mode III AFM.

*Optical Characterization*: PL measurements at room temperature and 77 K were conducted using a Horiba LabRAM system equipped with a confocal microscope, a charge-coupled device (CCD) Si detector, and a 532 nm diode-pumped solid-state (DPSS) laser as the excitation source. For temperature-dependent (above 77 K) measurements, the sample was placed into a microscope-compatible chamber with a low temperature controller (using liquid nitrogen as the coolant). In the experiment, the incident polarization angle was controlled by an angle-variable half-wave plate and was fixed, and the polarization angle of the emission (θ) was determined by using an angle-variable polarizer located in front of the detector. Time resolved PL measurements were conducted in a setup which incorporates μ-PL spectroscopy and a time-correlated single photon counting (TCSPC) system. A linearly polarized pulse laser (frequency doubled to 522 nm, with 300 fs pulse width and 20.8 MHz repetition rate) was directed to a high numerical aperture (NA= 0.7) objective (Nikon S Plan 60³). PL signal was collected by a grating spectrometer, thereby either recording the PL spectrum through a charge coupled device (CCD; Princeton Instruments, PIXIS) or detecting the PL intensity decay by a Si single-photon avalanche diode (SPAD) and the TCSPC (PicoHarp 300) system. The double exponential decay from 1L PEN is attributed to bi-exponential decay from instrument response function (IRF). The double exponential decay in the IRF is due to delay in response timings of the MPD SPAD® photon detector and the PicoHarp 300 coupled with it. A similar system response has been reported from an equivalent system.[62] Similar bi-exponential decay is commonly reported from such systems.[63] Thus, we have used deconvolution to extract a mono-exponential decay curve from 1L and all the results from lifetime data have been fitted after the deconvolution with the IRF (Figure 2c). This monoexponentially decay was further confirmed by our TR pump-probe femtosecond resolution measurements, which also clearly demonstrate a monoexponentially from 1L PEN at 77 K.

*Measurements of exciton diffusion lengths:* Exciton diffusion measurements were carried out using the same PIXIS CCD detector coupled with a X100 (NA= 1.49, oil suspended) objective lens. The pulsed 522 nm laser was used for excitation with a beam diameter ~500 nm (confirmed by CCD imaging) with collection time of 1s per measurement. The diffusion mapping direction was well aligned with the polarization of the excitation laser. The collected light was spectrally filtered to remove the pump laser wavelength. Spectral measurements were made using a grating spectrometer (Acton, SpectraPro 2750). The focal plane of the sample was imaged using the zeroth order of the grating and the spectrometer CCD, giving a spatial resolution of about 200 x 200 nm in space, corresponding to a pixel (20 μm x 20 μm) on the CCD. The PL intensity at a particular excitonic emission energy was plotted as a function of the distance from the excitation centre ($x = 0$ in Figures 3-4). The spatial extent of exciton diffusion (diffusion length $L_D$) was extracted by fitting the experimental data and laser profile with a 1D gaussian diffusion model [24, 44]. See Supporting Information Note 8.

*Pump-probe transient reflectivity measurements*: Transient reflectivity measurements were performed with a home-built pump-probe setup. Degenerate, linearly polarized pump and probe pulses were produced by splitting ~200 fs pulses from a non-collinear optical parametric amplifier (Light Conversion Orpheus-2N) tuned to 680 nm. A half-wave plate in the probe line rotated the polarization to be orthogonal to that of the pump. The pump was delayed using a motorized delay stage to control the pump-probe timing. Pump and probe pulses were recombined to be nearly spatially overlapped and focussed onto the sample with a 25.4 mm focal length lens. Telescopes in the pump and probe lines were used to tune the divergence/waist size of the pump and probe beams. The focal spot size and fluence of the pump and probe were 30 µm, 90 µJ/cm$^2$ and 15 µm, 12 µJ/cm$^2$ respectively.

The probe pulses were spectrally resolved across a fast CMOS array detector (Andor Zyla) by an imaging spectrometer (Andor Kymera). The detector recorded spectra at 5.21 kHz, synced to the laser repetition rate of 20.83 kHz. A data analysis technique based on lock-in detection was used to isolate only the component of the probe modulated at the same frequency as the chopper in the pump beam, which was also synced to the laser output. A polarizer was used to filter out the pump signal coupled with an iris. The residual pump still appears minimally both as a background signal and as interference fringes between the pump and the probe. The interferometric pump contributions were removed using a Fourier filter with a width of 400 fs, and the static background was removed by subtracting the signal observed at negative pump-probe delays (i.e. probe arriving before the pump). The data presented in Figure 2 is the result of 11 and 5 repeats of a 15 ps scan of the pump probe delay for 1L and WL respectively, with a step size of 100 fs and a 1 second dwell at each step.

*Spatial-temporal mapping*: Spatial-temporal mapping was carried out on our samples at 77 K using a movable lens, focusing the PL signal onto the CCD detector with a moving range of 10 µm from the central position on the sample (See schematic in Figure 4a). The mapping in the temporal decay from 1L and WL is consistent with the results shown in Figure 3 of the main text. The data shown in Figure S4b was smoothed in the post processing by using one level of interpolation (image spline) between the recorded data form the system to extract the diffusion coefficient due to limited number of data points in that regime.

**Author Contributions**

Y. L. conceived and supervised the project; M. D, L. Z, K. L. prepared pentacene samples; A. S. carried out all the optical measurements; A. S, Y. L. and R. H. analysed the data; A. S. took the AFM imaging; J.T., S.S., S.E., and J.D. helped with setup and measurements from TR pump-probe; H. T. N., D. M., T. V., P. K. L., Y. Z. and F. W. contributed to the set up for optical measurements; A. S. and Y. L. drafted the manuscript and all authors contributed to the manuscript.

**Supporting Information**

All additional data and supporting information and methods are presented in the supporting information file.

**Acknowledgements**

We would like to acknowledge the financial support from ANU PhD student scholarship, China Scholarship Council, Australian Research Council, ANU Major Equipment Committee fund, National Natural Science Foundation of China and Australian Centre for Advanced

Photovoltaics. We would also like to thank Professor Chennupati Jagadish and Professor Barry Luther-Davies from the Australian National University for their facility support. We acknowledge the helpful discussions with Professor Haibo Ma and Xinran Wang from Nanjing University, China.

**Competing financial interests**

The authors declare that they have no competing financial interests.

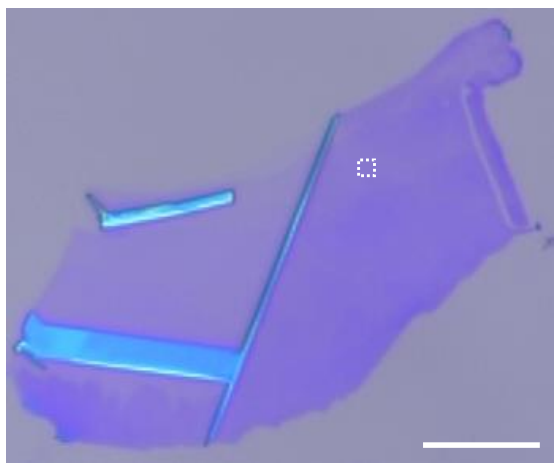
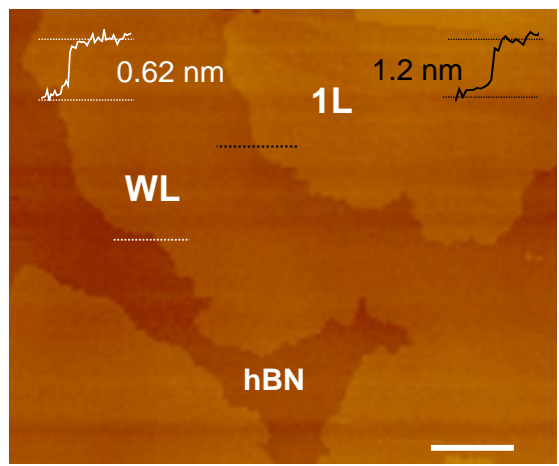
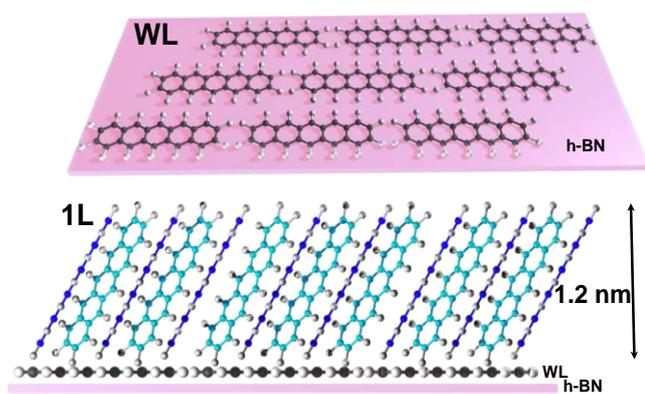
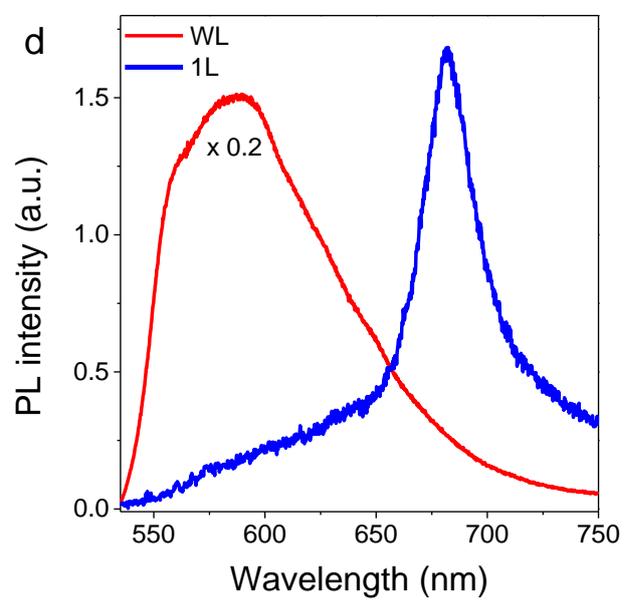

Figure 1

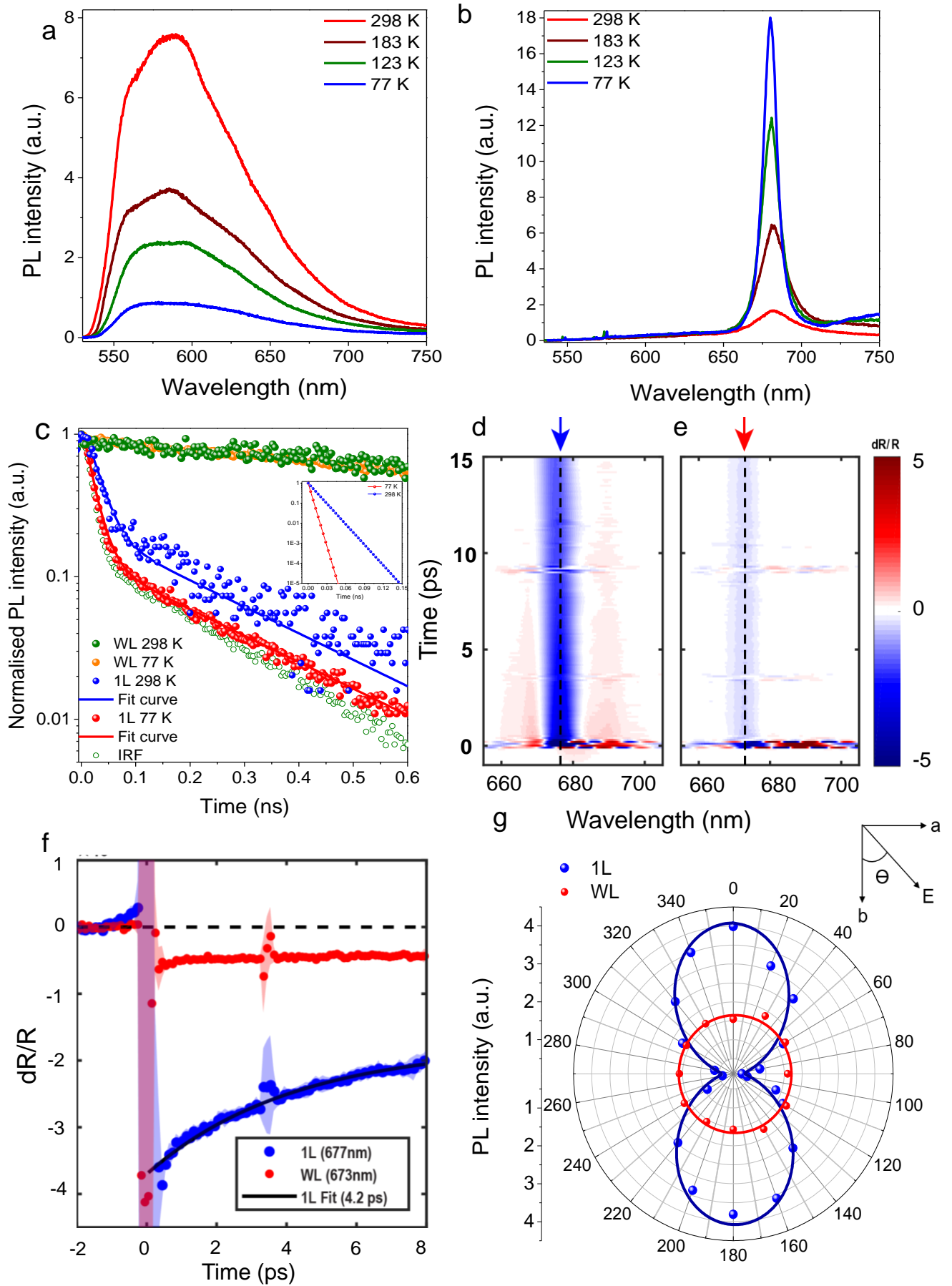

Figure 2

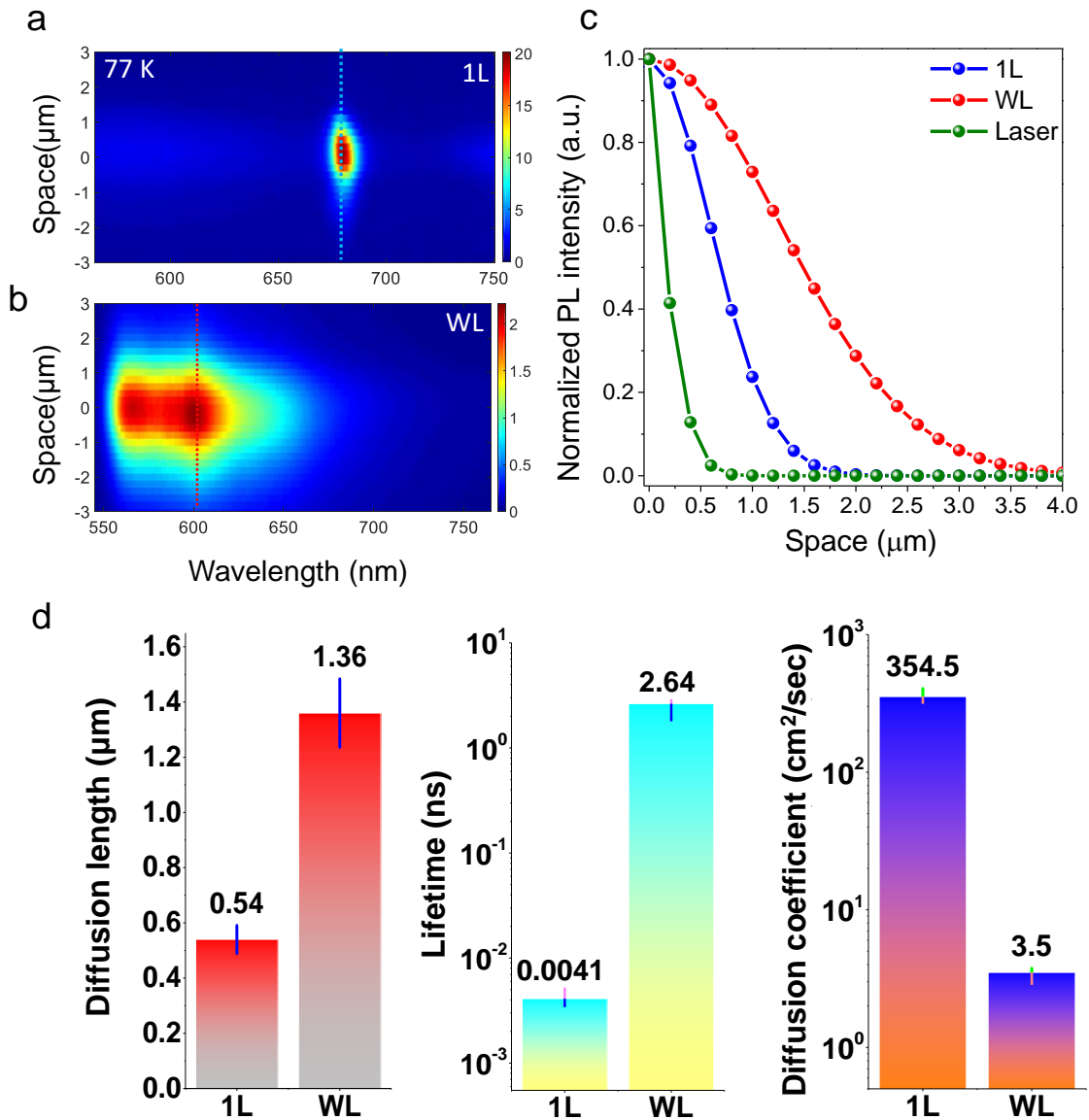

Figure 3

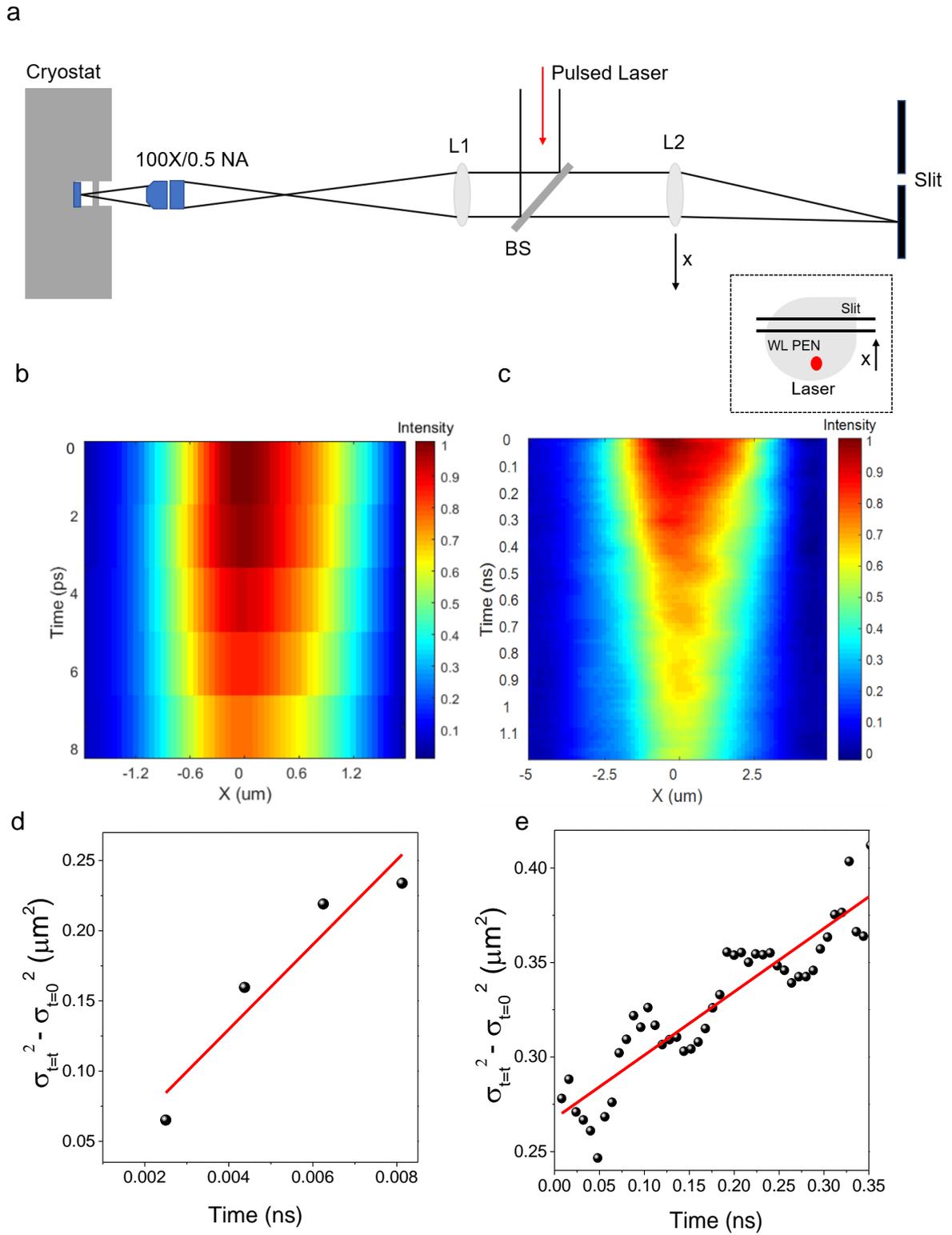

Figure 4

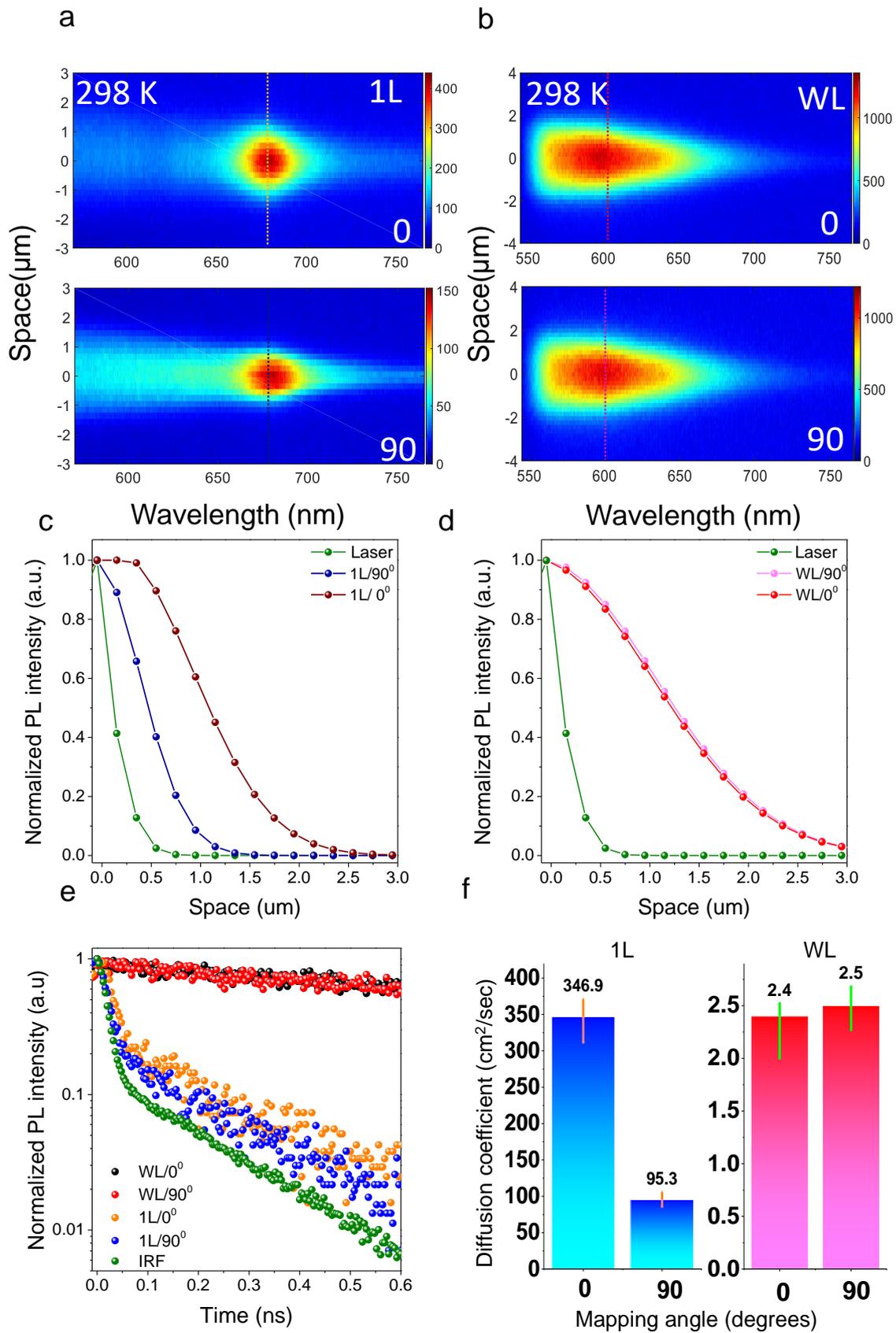

Figure 5

**FIGURE CAPTIONS**

**Figure 1 | Characterization of atomically thin 2D layered pentacene samples. a**, Optical microscope image of the sample used for measurements. The scale is 80 μm. **b**, Zoomed in Atomic force microscope (AFM) image of the dashed square region in (a), showing the actual measured thickness of WL and 1L pentacene. The scale is 2 μm. **c**, Schematic diagram showing the orientation and alignment of pentacene molecules in WL and 1L over h-BN and a $SiO_2$/Si substrate. **d**, Measured PL spectra from WL and 1L samples at room temperature.

**Figure 2 | Temperature-dependent optical measurements. a-b**, Measured PL spectra at various temperatures for WL (a) and 1L (b), respectively. For 1L, a sharp peak at ~680 nm (designated as FR excitonic emission) rises sharply as temperature decreases. **c**, Time-resolved PL emission (normalized) from WL and 1L pentacene samples at 77 and 298 K. The orange/green balls represent the measured decay curve from WL, which is from CT exciton emission. An effective long lifetime of 2.64 ns (2.89 ns) was extracted from the orange (green) decay curve at 77 K (298 K), by a fitting with deconvolution using the instrument response function (IRF) (green dots). The red/blue balls represent decay curve from 1L, which is from FR exciton emission. The red (blue) decay curve was fitted (solid line) with deconvolution using IRF, giving an effective short lifetime value of 4.1 ps (12.7 ps) for 1L at 77 K (298 K). The inset shows the deconvoluted curves obtained from 1L. **d-e**, Time and spectrally resolved transient differential reflectance of 1L (c) and WL (d) recorded using ~200 fs degenerate pump and probe with spectra centered at 680 nm. **f**, Slices of the TR for the 1L (blue) and WL (red) at 677nm and 673nm, respectively. A lifetime of 4.2 ps was extracted from the 1L decay, while the WL response was effectively constant across this range. The feature at 3.5 ps is due to interference between the pump pulse and a weak reflection of the probe pulse. **g**, Measured polar plot of PL intensity as a function of emission polarization angle $\theta$ from 1L (blue) and WL (red) pentacene samples at 77 K, revealing the anisotropic (isotropic) excitonic nature of emission from the 1L (WL) pentacene. In experiment, the excitation polarization angle was fixed and the polarization angle of the emission $\theta$ was determined by using an angle-variable polarizer located in front of the detector. The solid lines are fitted curves using a $cos^2\theta$ function.

**Figure 3 | Exciton transport mapping using near-field imaging. a-b**, Measured contour plots of PL intensity as a function of emission wavelength and space of exciton diffusion from 1L (a) and WL (b) pentacene samples at 77 K. The middle of laser excitation spot is at $x = 0$. **c,** Spatial profile plotted along the dotted lines in a-b for different types of excitonic peaks,

showing the diffusion lengths of excitons in WL and 1L. **d,** Comparison plots of diffusion lengths (left; from Figure 3a-c), lifetime (middle; from Figure 2e) and extracted diffusion coefficients, measured from WL and 1L samples, at 77 K. The error bars represent the experimental variation observed from measurements in multiple samples.

**Figure 4 | Spatial-temporal mapping of 1L and WL PEN, a,** Schematic showing the experimental setup with movable lens used for mapping. Pulsed 522 nm laser is focused onto the 1L/WL sitting on the copper cold finger in a cryostat using a beam splitter and 100 X/0.5 NA long working distance objective. Photoluminescence from the sample is imaged onto entrance slit of the spectrometer using lenses L1 and L2. Lens L2 is translated to position WL emission onto entrance slit of the spectrometer. **b (c)**, Spatial-temporal mapping from 1L (WL), showing the diffusion in time and space. **d (e)**, Time evolution of mean square displacement of excitons in space for 1L (WL). The obtained diffusion coefficient from 1L and WL is **306.8 ± 14.1** and **3.3 ± 1.1 cm$^2$/sec**, confirming the diffusion length as observed from Figure 3b.

**Figure 5 | Superfast and angle-dependent exciton transport at room temperature. a-b,** Measured contour plots of PL intensity as a function of emission wavelength and space from 1L (a) and WL (b) pentacene samples at room temperature, with spatial mapping direction parallel (top panel; marked as $0^0$) and perpendicular (bottom panel; marked as $90^0$) to the *b* axis of molecular lattices as shown in Figure S3a. In our measurement system, the mapping direction is parallel to the polarization angle of incident laser. The polarization angles $0^0$ and $90^0$ have been marked according to the references set in Figure 2f inset. **c-d**, Spatial profile plots along the dotted lines in a-b at two perpendicular mapping directions ($0^0$ and $90^0$), showing the angle-dependent exciton transport in 1L (c) and WL (d) pentacene samples. **e**, Time-resolved PL emission (normalized) from WL and 1L pentacene samples at room temperature. The black and red balls represent the measured decay curve from WL pentacene samples with incident polarization angle of $0^0$ and $90^0$, respectively. Effective lifetimes of 2.8 and 2.8 ns were extracted from the black and red decay curves, respectively, by fittings with deconvolution using the instrument response function (IRF) (green dots). The orange and blue balls represent the measured decay curve from 1L pentacene samples with incident polarization angle of $0^0$ and $90^0$, respectively. The orange and blue decay curves were fitted with deconvolution using IRF, giving an effective short lifetime value of 12.7 and 5.6 ps, respectively. **f**, Comparison plots of extracted diffusion coefficients measured from 1L (left panel) and WL (right panel) samples under the designated mapping angles at room temperature.

# Supporting Information

# Super-transport of Excitons in Atomically Thin Organic Semiconductors at the 2D Quantum Limit

Ankur Sharma,[1] Linglong Zhang,[1] Jonathan O. Tollerud[2], Miheng Dong,[1] Yi Zhu,[1] Robert Halbich,[1] Tobias Vogl,[3] Kun Liang,[1,4], Hieu T. Nguyen,[1] Fan Wang,[5] Shilpa Sanwlani[2], Stuart K. Earl[2], Daniel Macdonald,[1] Ping Koy Lam,[3] Jeff A. Davis[2] and Yuerui Lu[1*]

[1]Research School of Engineering, College of Engineering and Computer Science, The Australian National University, Canberra, ACT, 2601, Australia

[2]Centre for Quantum and Optical Science, Swinburne University of Technology, Hawthorn, Victoria, 3122, Australia and ARC Centre of Excellence for Future Low-Energy Electronics Technology, Swinburne University of Technology, Hawthorn, Victoria, 3122, Australia

[3]Centre for Quantum Computation and Communication Technology, Department of Quantum Science, Research School of Physics and Engineering, The Australian National University, Acton ACT, 2601, Australia

[4]School of Mechatronical Engineering, Beijing Institute of Technology, Beijing 100081, China

[5]Institute for Biomedical Materials and Devices (IBMD), Faculty of Science, University of Technology Sydney, NSW 2007, Australia.

\* To whom correspondence should be addressed: Yuerui Lu (<u>yuerui.lu@anu.edu.au</u>)

**Supporting Information Note 1: Exciton types and theory**

Excitons are defined as the quasi bound state of electron and hole pair, which are attracted to each other by coulombic electrostatic forces. Based on the degree of charge separation which in turn is determined by material dielectric, the excitons can be broadly classified into the following categories[1]:

*Wannier-Mott* (W-MT) excitons are generally found in materials with large dielectrics, thus electric field screening reduces the Coulomb interaction between $e^-$ and $h^+$. This results in much larger size for this kind of excitons, which is sometimes larger than space lattice. In W-MT excitons, smaller effective mass of $e^-$ also favours the large radii of excitons. Typical binding energies reported here are around 0.01 eV. They are found mostly in inorganic semiconductors have been recently shown to exist in monolayer TMDs as well.

*Frenkel (FR)* excitons, whereas are found in materials with low dielectric constants, the Coulomb interaction between $e^-$ and $h^+$ is much stronger, thus have smaller radii. They are observed mostly in organic materials such as anthracene, pentacene and other materials with low dielectric constants, thus enhancing the Coulomb interaction between $e^-$ and $h^+$ resulting in much smaller radii. The size is of the same order as a unit cell and often reside on the same molecule. Binding energies range from 0.5-1.0 eV.

*Charge transfer* (CT) excitons are the intermediates between FR and W-MT excitons. They occur when $e^-$ and $h^+$ occupy adjacent molecules. They are considered as a localized, structured version of W-MT exciton, where a higher dielectric constant permits the radius to expand over many sites. Their excitonic wave function resembles that of a hydrogen atom. Both CT and FR excitons can have localized wave functions.

**Supporting Information Note 2: Role of molecular packing and temperature on optical properties**

The effect of molecular packing on excitonic states was firstly reported by Kasha *et al.*[2], who extended the geometry-related FR exciton treatment by Davydov.[3] In our model, we define $J$ to be excitonic coupling between different local FR exciton states located at nearest-neighbour molecules. The oscillator strength of an optical transition is a function of the total TDMs in an aggregate of molecules as shown below theoretically. If the Coulombic repulsion is stronger than the Coulombic attraction, their competition will favour a positive $J$ value ($J > 0$) and vanishingly small total TDM, which will lead to an optically forbidden lowest excited state ($S_1$) (Figure S3b)[4]. In contrast, if the Coulombic attraction is stronger than the Coulombic repulsion, their competition will favour a negative $J$ value ($J < 0$) and large total TDM, which will result in an optically allowed $S_1$ (Figure S3c). The result of this competition mainly depends on the orientation rather than the distance between molecules.[5] Molecular aggregates are usually considered to be H-type when $J > 0$ and J-type when $J < 0$.[5] Generally, side-by-side molecular orientations lead to $J > 0$, whereas head-to-tail orientations lead to $J < 0$.

For a transition from state $|\Psi_a\rangle$ to another state $|\Psi_b\rangle$, its transition dipole moment (TDM) is[6]

$$\text{TDM}(a \rightarrow b) = \langle \Psi_a | \hat{\mu} | \Psi_b \rangle = \langle \Psi_a | q\boldsymbol{r} | \Psi_b \rangle = \int \Psi_a^*(\boldsymbol{r}) q(\boldsymbol{r}) \boldsymbol{r} \Psi_b(\boldsymbol{r}) d\boldsymbol{r} \qquad (1)$$

where $q(r)$ is the charge on the particle at an arbitrary position '$r$'. TDM is not the oscillator strength *(f)*, but it is the most crucial parameter for determining $f$.[7]

$$f(a \rightarrow b) = \frac{2m_e(E_b - E_a)}{3\hbar^2} \text{TDM}(a \rightarrow b)^2 \qquad (2)$$

where, $E_b$ and $E_a$ are the energy levels associated with states '*b*' and '*a*' respectively.

In short, the magnitude of TDM determines the strength of the optical transition to a large extent. Additionally, its $\alpha$ (= $x, y, z$ or $a, b, c$) component determines the incident light polarized absorption (or transition) along $\alpha$ direction.

**Supporting Information Note 3: Coherence number determination and effect of vibration and phononic interaction**

The high crystallinity obtained in 1L pentacene samples will result in large coherence lengths and high diffusion coefficients but the effect of vibrations or phonon interaction in the lattice cannot be ignored. The phonon or vibrational coupling affects the PL spectra which is characterized by a spectral phononic sideband emission in addition to the sharp 0-0 Frank Condon emission, which is at 680 nm in our case (See Main Text Fig 2). As suggested by Zhao *et al.*[8] and Tanaka *et al.*[9], the coherence number will start to decrease as the vibrations increase and hence resulting in lower diffusion lengths and diffusion coefficients.

To quantify the phonon interaction we have followed the technique reported by Spano *et al.*[10] to precisely determine the coherence number ($N_c$) in 1L pentacene at 298 K and 77 K. In effect, this vibration serves as a probe for the coherence length. When the optical transition from the lowest exciton to the vibration less electronic ground state (the "0−0" transition) is symmetry-allowed, as in a J-aggregate, the ratio of the 0−0 to 0−1 transition line strengths can be used to determine the coherence number. The coherence number can be very precisely number determined taking into account the spectral line strengths (integrated PL) of the coherent peak and the sideband emissions by the following equation:

$$\frac{N_c}{\lambda^2} = \frac{I_{0-0}}{I_{0-1}}$$

Here, the $I_{0-0}$ is referred to the FR emission at 680 nm and $I_{0-1}$ is the phonon sideband emission caused due to phonon or vibrations in the lattice. $N_c$ is the coherence number and $\lambda^2$ is the Huang-Rhys factor in pentacene. At 298 K and 77 K the PL spectra can be fit by two Lorentzian peaks as shown in Figure S6a and S6b respectively.

The Huang-Rhys factor used for our calculation is taken to be 0.574 which has been theoretically determined for pentacene and taken from ref[11]. The extracted values of spectral strengths and 0-0 and 0-1 transitions and subsequent $N_c$ has been tabulated as below.

| Temperature (K) | $I_{0-0}$ (a.u.) | $I_{0-1}$ (a.u.) | $N_c$ |
| --- | --- | --- | --- |
| 298 | 84858 | 1615 | 30 |
| 77 | 287336 | 1214 | 135 |

The value of $N_c$ derived at low temperature is much higher than previously reported values from thin film pentacene and similar oligoacene films like tetracene ($N_c$= 10), thus confirming the limited effect of vibrational or phononic coupling in high crystalline 1L PEN samples specially at low temperatures. The high Value of $N_c$ directly corresponds to higher diffusion lengths and higher diffusion coefficients in 1L PEN. The value of coherence number reported is an order of magnitude higher than previously reported values from thin film organic semiconductors and hence a high value of diffusion coefficient is expected from such a system operating with large coherent regimes to allow for excitons to transport quickly.

**Supporting Information Note 4: Excitonic character in WL and 1L governed by molecular packing**

The excitonic nature of organic crystals plays an important role in determining their optoelectronic properties.[4,10,12] In organic semiconductors FR and CT excitons are most commonly observed.[13,14] (See Supporting Information Note 1). As soon as the energy difference between CT and FR excitons becomes small, both excitons can interact and form new FR–CT mixed states.[14] Intermolecular coupling also influences the complex excitonic character in organic molecules. The mixed states are observed in most bulk and thin film organic semiconductors due to the lack of high crystallinity, scattering and the presence of disorders and interfacial states.[15-17] Precise and controlled growth of organic molecular crystals in atomically thin 2D states provides an ideal platform for isolating the smaller and tightly bound FR[18] excitons as they have closer packing of molecules within a unit cell.[18-20] In order to reduce the FR-CT mixing and to observe the aggregation-controlled coherence phase locking and subsequent superradiant emission, the crystallinity and molecular aggregation into H- (CT exciton dominated aggregation) and J-type (FR exciton dominated molecular aggregation) has to be controlled, as theoretically predicted by Spano *et al*.[10]

In our case, the highly clean and sharp crystalline system allowed us to observe long-range coherent super-transport from FR excitons which are de-localized over hundreds of pentacene molecules, due to of the minimized defect, disorders and interfacial states in the monolayer regime. WL pentacene samples, which had H-type molecular aggregation, showed in-plane isotropic emissions that were dominated by CT excitons. In contrast, 1L pentacene samples, which had J-type molecular aggregation, showed one sharp and strongly anisotropic PL emission peak (See Supporting Information Note 2 for theoretical simulation details). The WL region showed much stronger PL emissions than the 1L regions at room temperature (Figure 1d). Compared with inorganic TMD 2D semiconductors, the WL shows a much broader PL spectra at room temperature, which is due to the strong couplings between FR and CT excitons in organic materials.[18,21] The broad PL spectrum is associated with various band energy levels formed in pentacene due to vibronic coupling between FR and CT states[10]. 1L on the other hand shows a single peak centred around 680 nm which is predominantly FR excitonic emission as explained by our polarization dependent measurements in Figure 2g and also explained below. We grew several samples of pentacene in various growth conditions, giving us different optical properties from 1L depending on growth conditions (Figure S3). The poorly grown 1L samples also gave us broad high energy CT excitonic peaks in addition to the narrow FR excitonic peak at 680 nm overlapping with the emission energies from WL in certain samples, showing the mixed FR and CT emissions from 1L. Whereas, some 1L samples gave us only the CT emission with little FR emission. This clearly highlights the role of well-defined molecular arrangements in de-coupling the intermixing of FR and CT excitons. We have chosen the 1L sample which gave us only FR emission for further analysis to independently study the behaviour of CT and FR excitons in WL and 1L *viz*.

We also performed PL measurements with an angle-resolved emission polarization (Figure 2f) to confirm the anisotropic nature of superradiant FR excitonic emissions from 1L. In the experiment, the incident polarization angle was controlled by an angle-variable half-wave plate and was fixed, and the polarization angle of the emission ($\theta$) was determined by using an angle-variable polarizer located in front of the detector. In polarization-dependent PL measurement at 77 K, we found that intensity of PL emission from 1L strongly depended on the emission polarization angle $\theta$ and shows a period of 90 degrees. On the contrary the emission from WL was completely anisotropic (Figure 2f). The strongly directional total TDM in 1L pentacene predicted by our simulation (Figure S3a) was expected to lead to an in-plane anisotropic emission of those FR excitons. The observed strong anisotropy of PL emission from 1L pentacene further confirms the strong optical alignment of molecules in a specific orientation and agrees very well with the prediction from our simulations, which confirms that the superradiant emission from 1L pentacene observed in 1L and 2L pentacene at low temperature is dominated by FR excitons.

**Supporting Information Note 5: Coherent exciton transport and supertransport**

The diffusion of FR excitons that are commonly found in organic semiconductors migrate much slower as compared to their Wannier counterparts in inorganic materials. This is because of lower dielectric constant in organic materials and results in lower diffusion lengths in organic materials.[22,23] The energy transfer is dominated by incoherent Förster resonant energy transfer mechanism (FRET)[24], which involved hopping of charges from one chromophore to the other *via* dipole moment interactions. This mechanism is based on dipole-dipole electromagnetic interaction and occurs when emission spectrum of donor molecule has significant overlap with absorption of the acceptor molecule. It is a non-radiative transfer and relies on long-range Columbic coupling between the nearby molecules.[23] Due to this incoherent hopping, the diffusion of FR excitons is limited. Incoherent energy transfer is generally determined my inter-molecular spacing, defect states, excitonic lifetime and range of columbic coupling also referred to as Förster radius.[22,25] All these factors lead to smaller diffusion of FR excitons inorganic molecules and molecular assemblies.

However, in organic molecules the inter-molecular coupling can become very strong if the crystallinity is strong.[26,27] If the coupling between organic molecules is strong, the FR excitons can be delocalized over multiple molecules forming a delocalized coherent state for the exciton, which eventually leads to superradiant emissions.[28] In this phenomenon interactions between transition dipoles of individual molecules allow coherent delocalization across multiple sites. The corelated states lead to a macroscopic optical polarization proportional to the number of atoms or molecules comprising the coherent domain.[29,30] This leads to a net enhancement in the optical transitional dipole moment (TDM) value and sharp enhancement of the excitonic radiative decay rate of an ensemble of $N_c$ independent emitters as compared to the radiation decay rate of a single emitter.[31,32] The same principle of coherent delocalized superradiant emission gives rise to an analogous phenomenon called cooperative energy transfer or supertransfer (ST).[33] The resulting enhanced oscillator strength from delocalization over large molecular assemblies can lead to large scale exciton transport. Here, the molecular assemblies

(consisting of $N_c$ molecules) with comparable net TDMs can play the role of acceptors like individual molecules in FRET mechanism.[24] Thus, the excitation (exciton) can transfer to much longer distances in a coherent delocalized molecular assembly before annihilation, resulting in large values of diffusion coefficients. This strong intermolecular coupling plays an important role in developing coherent effects and determines long-range coherent exciton transport, which has been demonstrated in some quasi 1D systems recently. Delocalized excitons principally accelerate energy transfer as compared to incoherent hopping transfer because delocalization allows for a hopping length that is much larger as compared to intermolecular spacing.[25]

**Supporting Information Note 6: Relationship between crystalline order and diffusion length**

As discussed above the charge or energy transfer mechanism in organic molecules requires hopping of excitation from one molecule to other (incoherent) or from one delocalized state to other (coherent). In either case, the distance between interacting molecules or states is critical. This intermolecular distance is a function of degree of crystallinity in the molecule.[34] Hence, the diffusion of excitons in organic system is a direct function of crystallinity. Lunt *et al*.[35] and Sim *et al*.[34] recently experimentally demonstrated that exciton transport diffusion length is a monotonic function of the extent of crystalline order. The reduction in diffusion lengths with decreasing crystallinity is due to increase in non-radiative losses in highly disorders systems, as determined by fluorescence quantum yield measurements as a function of grain size in Perylenetetracarboxylic dianhydride (PTCDA).[36]

We observed a similar trend during the growth optimization process of our pentacene samples. The samples which used h-BN surfaces that were exposed to air for a long time or had uneven morphology (detected optically) gave us low crystalline growth, as characterized by AFM measurements. In contrast the samples grown on homogenous and unexposed h-BN substrates gave a much stronger crystallinity and smooth pentacene growth, confirmed by surface roughness measurements. All the samples were characterized by AFM right after their growth. During the growth of pentacene samples in our case, we observed that the optical performance was also critically affected by the order of crystallinity in our growth as shown in Figure S5. The samples that had lower crystallinity showed a PL spectrum similar to earlier reports.[37] The broad PL spectrum obtained even from 1L samples is due to vibronic mixing of CT and FR states (Samples 2-4 in Figure S5). After several rounds of optimization, we could precisely control the growth conditions and orientation of molecules in 1L (Sample 1 in Figure S5). There were no CT peaks reported from highly crystalline samples, demonstrating only a sharp FR excitonic emission at 680 nm. Due to reduced defects, lack of interfacial sites and high crystallinity in our samples we could observe higher diffusion coefficients even for short-lived FR excitonic emissions from 1L pentacene samples.

**Supporting Information 7: Dimensionality and oscillator strength**

Apart from coherent delocalization and high order crystallinity, the exciton diffusion process in organic molecules is also a function of the oscillator strength of the excitons.[23,26,27,38] The oscillator strength in return is strongly influenced by the effective dimensionality of the solid-state system as theoretically demonstrated by Wu *et al*.[39] The oscillator strength can be highly enhanced by reducing the effective dimensionality of the system. Thus, it is critical for us to develop a true 2D system that can confine the excitons in 2D quantum limit to further enhance the diffusion of the FR excitons.

The quantum confinement effect plays a key role in determining the spatial diffusion of excitons.[27] For the same reason, several recent attempts to achieve higher diffusion coefficients with molecular aggregates have used quasi-1D structures like tubes[25] or wires[27] to enhance the diffusion using J-type aggregation. In our pentacene WL and 1L system, the excitons were spatially confined in a 2D space, which increases the possibility of directed excitonic energy transfer within the 2D plane with limited out of plane losses and hopping. Further, in 1L as we have established the excitons were found to be anisotropic and aligned along the '*b*' axis of the unit cell of pentacene (Figure 2f in main text and associated text). This led to further confinement and transport if excitons are in a quasi-1D system, which eventually led to observation of ultra-high exciton diffusion coefficients in our samples.

**Supporting Information Note 8: Diffusion Length using direct CCD imaging**

A robust method of determining exciton diffusion is in steady state which involves modelling of the spread of excitons in a 1-D model. In the limit of low exciton density, the number of excitons n (x, t) as a function of both time and position can be described from the diffusion equation.

$$\frac{\partial n(x, t)}{\partial t} = \frac{n(x, t)}{-\tau} + D\frac{\partial^2 n(x, t)}{\partial x^2} + pI(x)$$

where, D is the coefficient of diffusion, is the exciton lifetime measured separately, p is the probability of absorption and I(x) is the spatial profile of excitation. If we assume a gaussian profile with a standard deviation of σ (or full width half maximum-FWHM), the solution will lead to the following equation:

$$n(x, t) \propto \frac{1}{(2\sigma_0^2 + 4Dt)^{1/2}} \exp\left(\frac{-x^2}{(2\sigma_0^2 + 4Dt)}\right) \exp\left(\frac{-t}{\tau}\right)$$

In our measurements in Figure 3 and 4, the profile measured is a steady state profile unlike in measurements in Figure S12 where is a function of time. This steady-state profile is modelled by numerically integrating the above equation over time. This modelled profile is the used to fit the experimental data from Figure 3a, b to extract the diffusion length or s. We have then also fit the excitation laser gaussian profile using the same fitting function to extract its width.

Both emission and laser profiles were fitted using the 1D Gaussian distribution model and their FWHMs were extracted. The spatial extent of exciton diffusion (diffusion length $L_D$) was extracted by taking the difference of these extracted widths from the fitting.

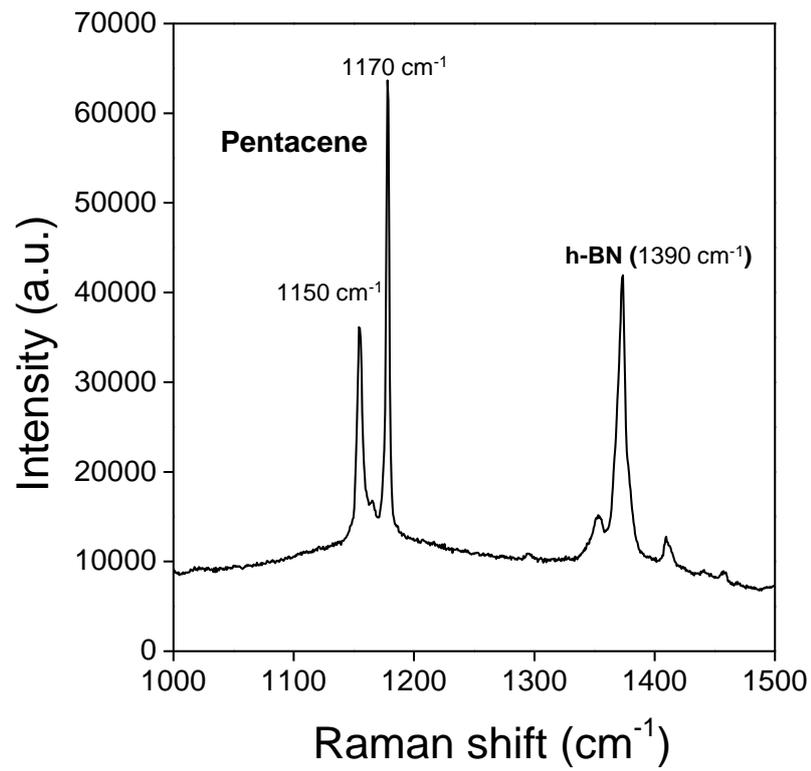

**Figure S1** | Measured Raman spectrum from a 2D pentacene sample grown on a h-BN flake. Peaks at 1150 cm$^{-1}$ and 1170 cm$^{-1}$ are for pentacene and peak at 1390 cm$^{-1}$ is for h-BN.

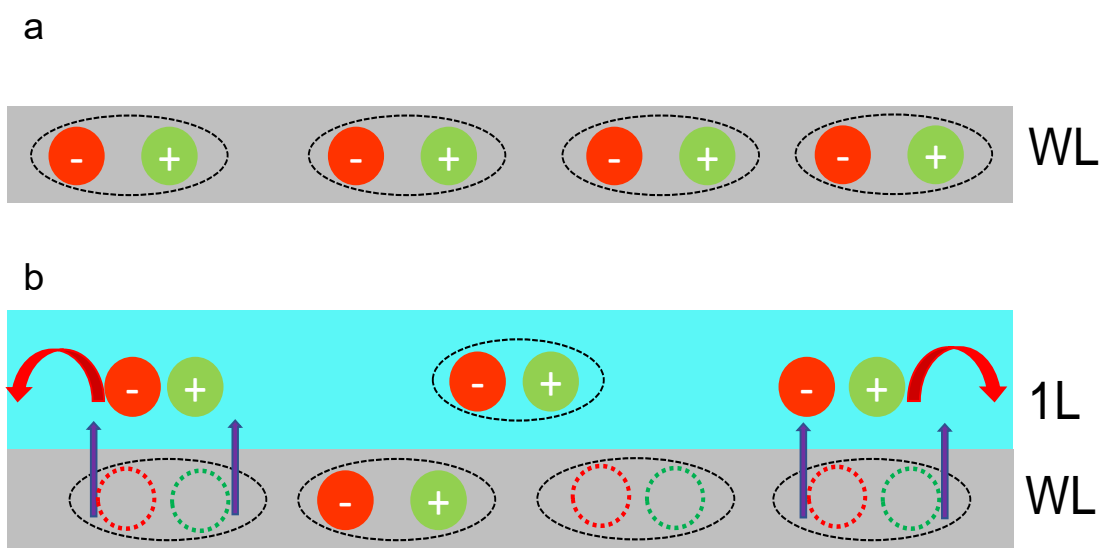

**Figure S2 |** Schematic diagram showing the interlayer charge transfer and different intra-layer charge transport mechanism in WL (a) and 1L+WL (b). As shown in a previous report[37], WL pentacene is not conducting due to the absence of intralayer π-π stacking. The charges interact by long range coulombic coupling in WL; in contrast 1L shows 2D hopping transport. The purple arrows represent interlayer charge transfer. The red curved arrows represent the hopping of charges in 1L. Here, we only show the photo-excited charges generated in WL, not in 1L. The charges interacted by long-range Coulombic coupling in WL (Figure S2a); in contrast, 1L showed 2D hopping transport. The photo-excited charges in the 1L layers had much higher in-plane diffusion than those in WL, which led to lower PL efficiencies in 1L pentacene. For the 1L+WL hetero-structure (1L region), the photo-excited charges in WL underneath could be quickly transferred to 1L and diffused laterally in 1L before radiative emission occurred (Figure S2b). This led to quenching of the PL intensity. This type of ultrafast interlayer charge transfer phenomena has been observed for both inorganic 2D hetero-structures[40] and organic–inorganic interfaces.[41]

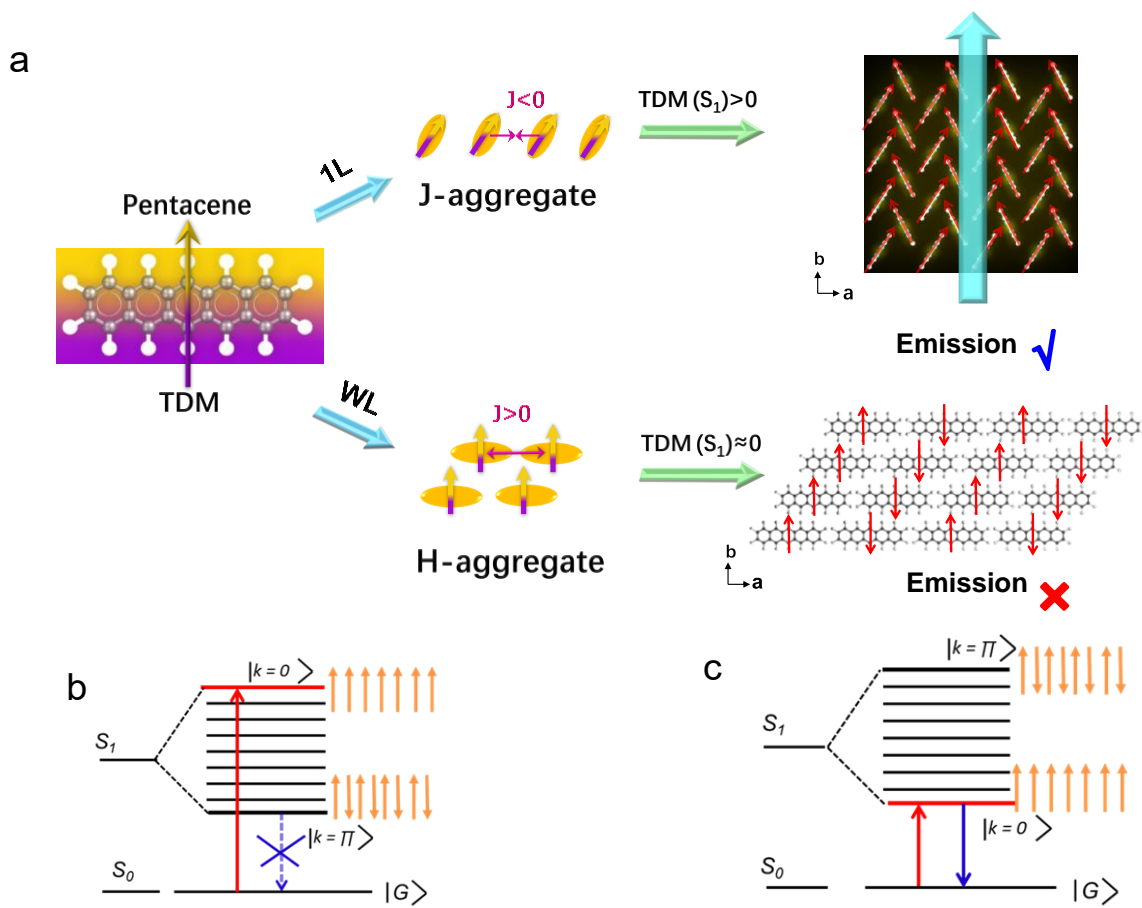

**Figure S3 | Molecular packing and the FR exciton model. a,** The TDMs at each molecule (red arrows) follow a head-to-tail packing in 1L as shown, while in WL TDMs are packed side-by-side to each other. The pink arrows represent the Coulombic interaction force between neighboring molecules. $J$ is the excitonic coupling between different local FR exciton states located at nearest-neighbor molecules. Based on our simulation, $J$ is negative in 1L (corresponding to J-type aggregation) and hence there is a net positive total TDM value along the "$b$" axis (as shown by purple arrow) allowing for emission from 1L, while $J$ is positive in WL (corresponding to H-type aggregation) and hence a close to null total TDM, resulting in very little emission from WL. **b, c** Schematics showing the excited states in an ideal H- (a) and J-type (b) monomer. Optical transitions from and to ground state are only allowed for excited state band with $k = 0$, where $k$ is the quantum number of the wave vector, representing the excitonic states. Orange arrows indicate the spin of electrons in each state. In J-type the 0-0 transition results in enhanced coherent emission due to coupling of TDMs, eventually causing super radiant emissions.

**Reason for blue-shifted WL PL spectra:** The WL is blue-shifted because it is an H-aggregate as compared with the spectra from 1L/2L. This can be understood by looking at the schematic in Figure S3b (See below), where the optically allowed bands are always at much higher energies as compared to the J-type aggregates. This blue-shift from H-aggregates was also predicted by Kasha *et al.*[2] and Spano *et al.*[9]

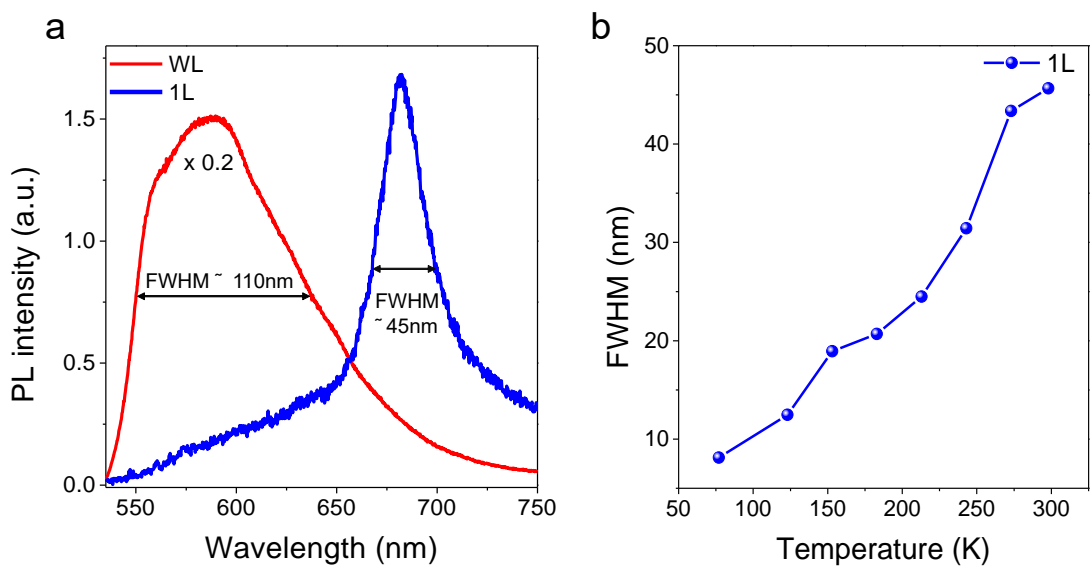

**Figure S4 | Measured FWHM with decreasing temperature. a**, PL spectra and respective FWHM at 298 K from WL and 1L. **b,** Measured Full-width half maximum (FWHM) as a function of temperature from 1L PEN.

The FWHM from WL overall is much higher at 298 K as compared with narrow linewidth from 1L, suggesting a dominant coherent component in the emission. As shown in the figure above the WL spectra is very board with FWHM ~110 nm, which arises due to CT and FR mixing in the WL. The FWHM from 1L has been plotted as a comparison show the extent of reduction in 1L as compared to broad WL PL spectra.

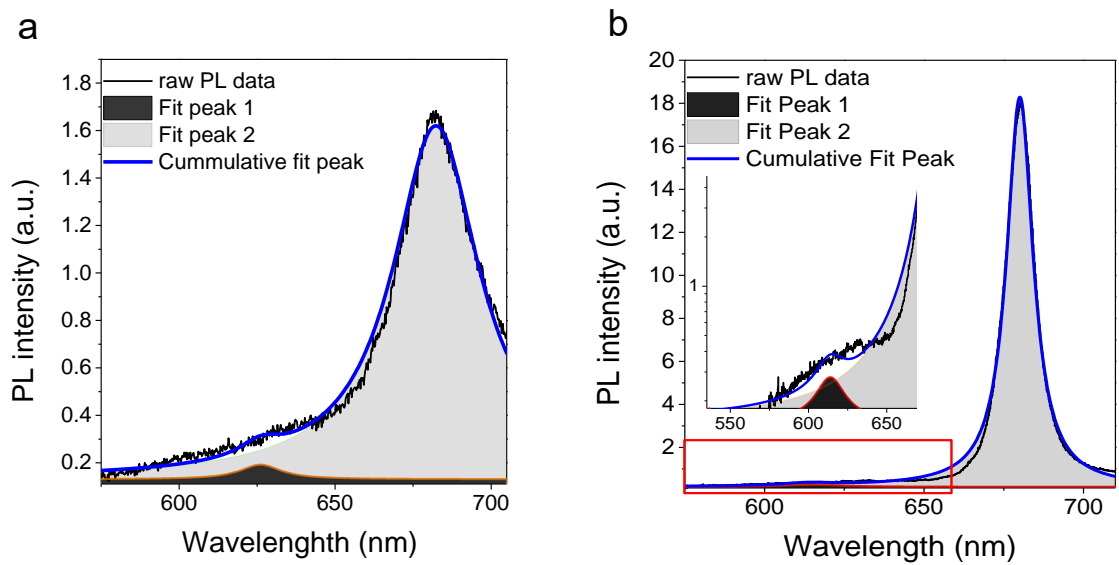

**Figure S5 |** PL spectra line strengths from 1L PEN. **a**, PL spectra as measured at 298 K. The shaded areas under the curve after a Lorentzian fitting function have been used to fit the PL data. **b**, PL spectra as measured from 1L PEN at 77 K fitted using two Lorentzian peaks. The inset shows a zoomed in area marked in red square on a log scale.

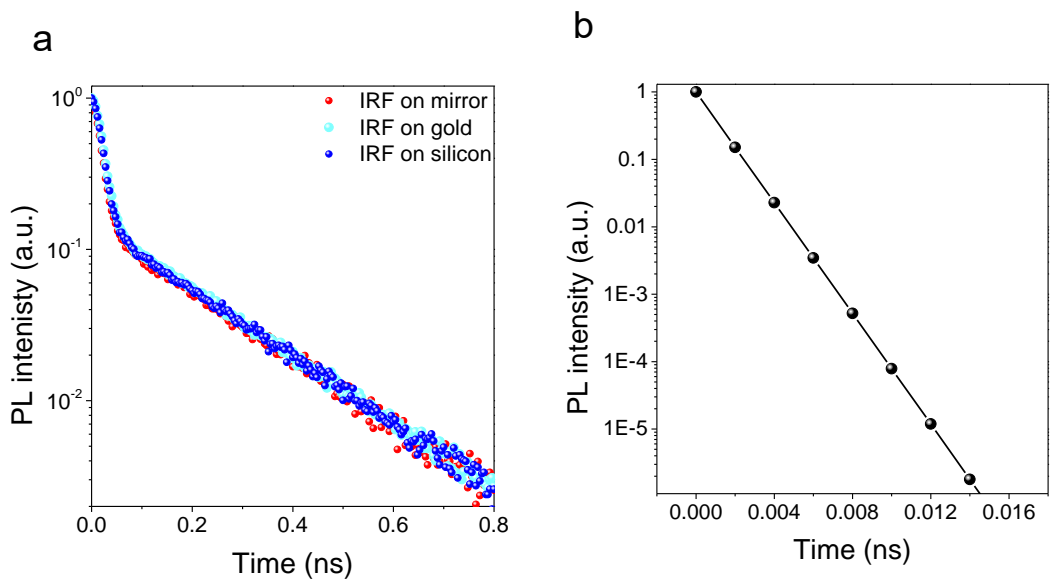

**Figure S6 |** Instrument response functions. **a**, as measured from different reflective surfaces. The IRF is very stable from different surfaces. **b**, Deconvoluting of one IRF over other gives a system resolution of 2.1 ps as shown by the deconvoluted graph.

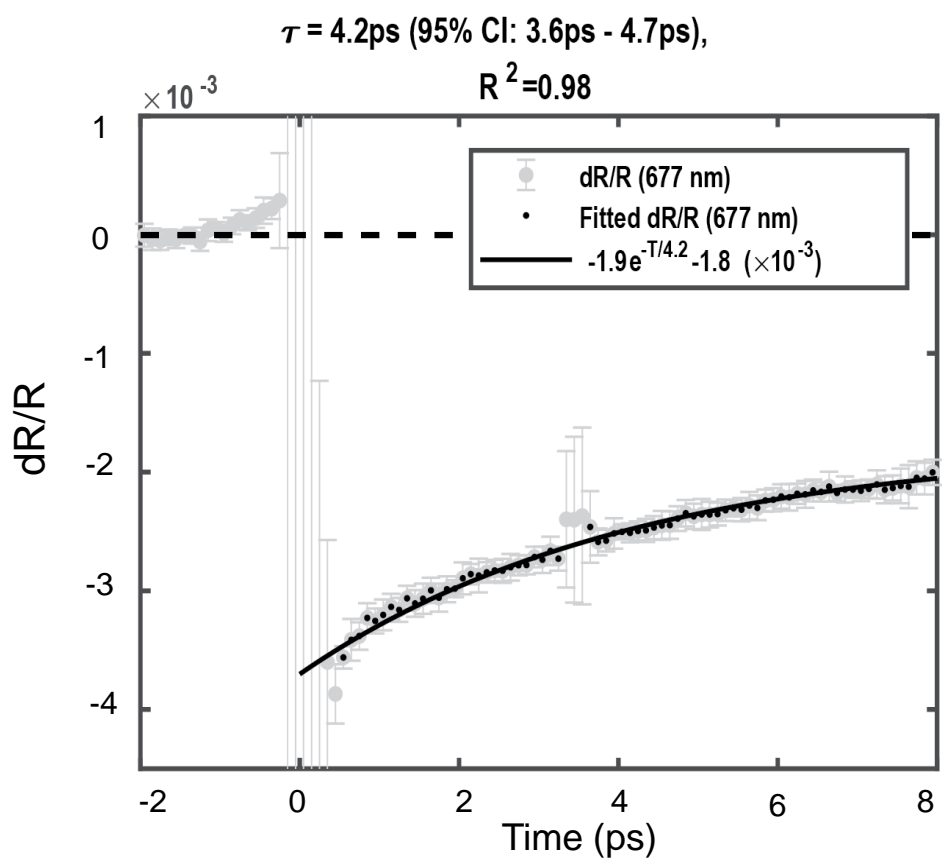

**Figure S7 | Fitting and error analysis.** The figure shows the fitting of the decay curve from pump-probe measurements with a 95% confidence interval fiving a lifetime range of 3.55-4.6 ps and the best fit with a value of 4.2 ps.

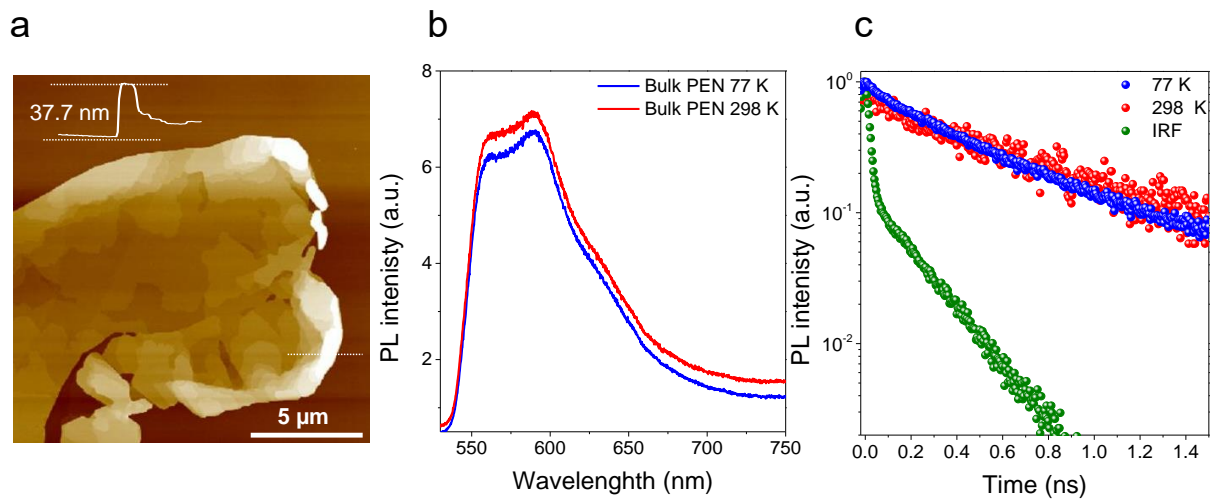

**Figure S8 | PL emission from bulk thin-film pentacene, a,** AFM image of the bulk thin film pentacene grown over h-BN with a thickness of around 37.7 nm. **b**, PL emission at 77 K from the bulk pentacene, which is dark as compared to WL. **c**, the decay curve from bulk pentacene at 298/77 K, giving an effective lifetime of 1.241/1.014 ns.

The emission from bulk pentacene layers is a broad PL spectrum, which changes negligibly with lowering of the temperature. The sharp peak emission at 680 nm is missing from the bulk pentacene. The broad PL spectra can be attributed to the coupling of CT and FR states, arising from the disorders and interfacial states, which has been reported in literature previously as well.[10,12] The lifetime from the bulk pentacene peak at 600 nm emission is 1.241 ns at 298 K, which changes to 1.014 ns at 77 K. The linewidth of the spectra is one order of magnitude higher (>100 nm as compared to 8 nm from 1L pentacene) from bulk pentacene, suggesting a non-coherent emission.

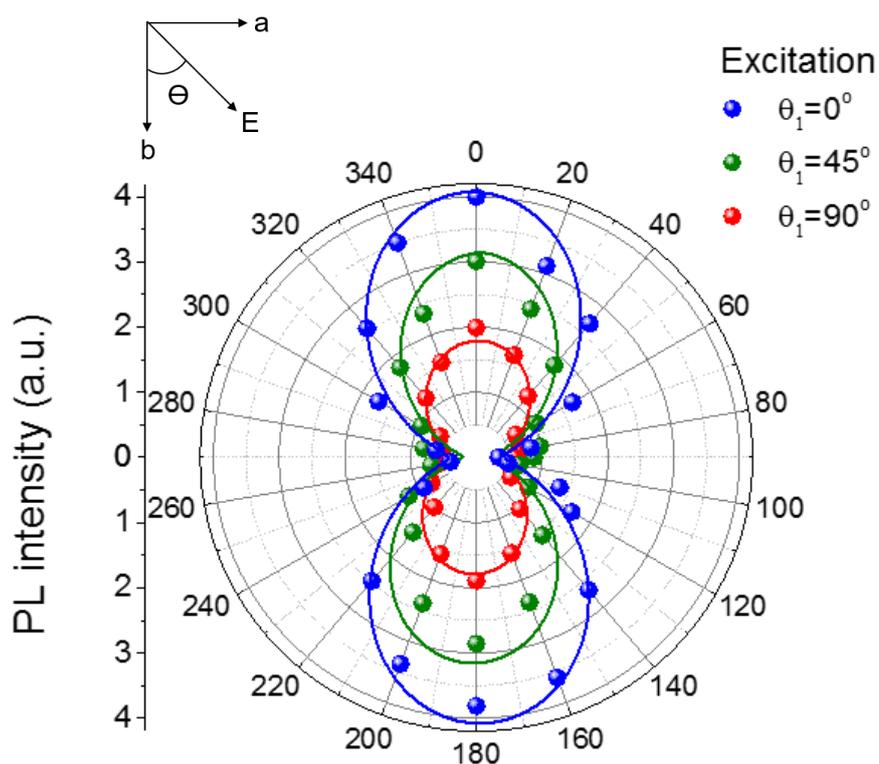

**Figure S9 | Emission polarization measurements from 1L PEN at different laser polarizations.** The polarization dependent PL measurements from 1L at 77 K. In experiment, the excitation polarization angle $\theta_1$ was fixed at 0° and the polarization angle of the emission ($\theta_2$) was determined by using an angle-variable polarizer located in front of the detector. We also changed $\theta_1$ and repeated the measurements. We observed that PL emission from 1L always shows the maximum PL intensity at $\theta_2 = 0°$ and the minimum PL intensity at $\theta_2 = 90°$, regardless of the excitation polarization angle θ1 used 0° (grey), 45° (magenta) and 90° (blue).

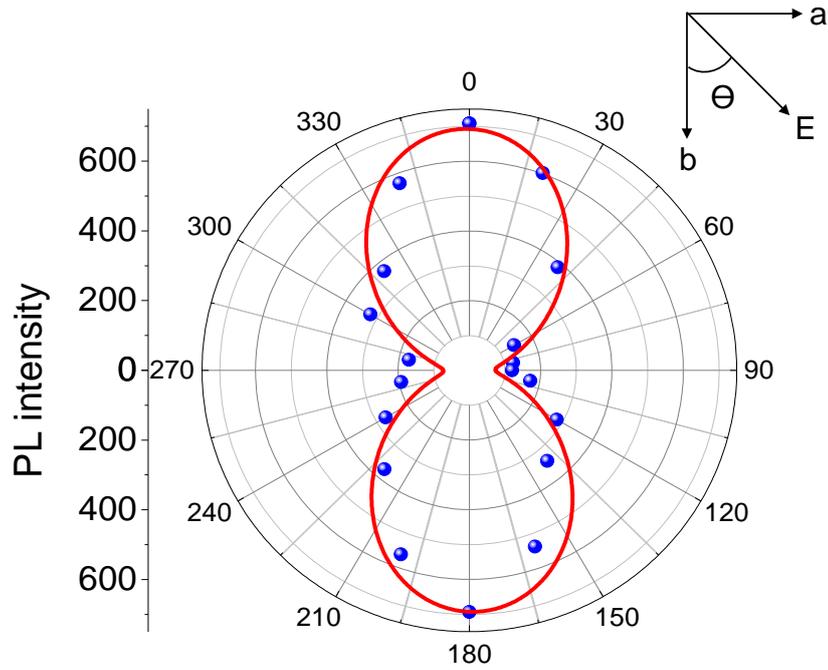

**Figure S10 | Measured angle-resolved PL for excitation polarization from the 1L at room temperature. a**, Measured angle-resolved PL spectra for emission polarization from 1L pentacene. The LDR is around 5.9 at room temperature. The incident polarization angle ($\theta$) was controlled by an angle-variable polarizer. The excitation laser power remained constant. The solid balls represent the measured experimental values and solid lines are fitted curves using a $cos^2\theta$ function.

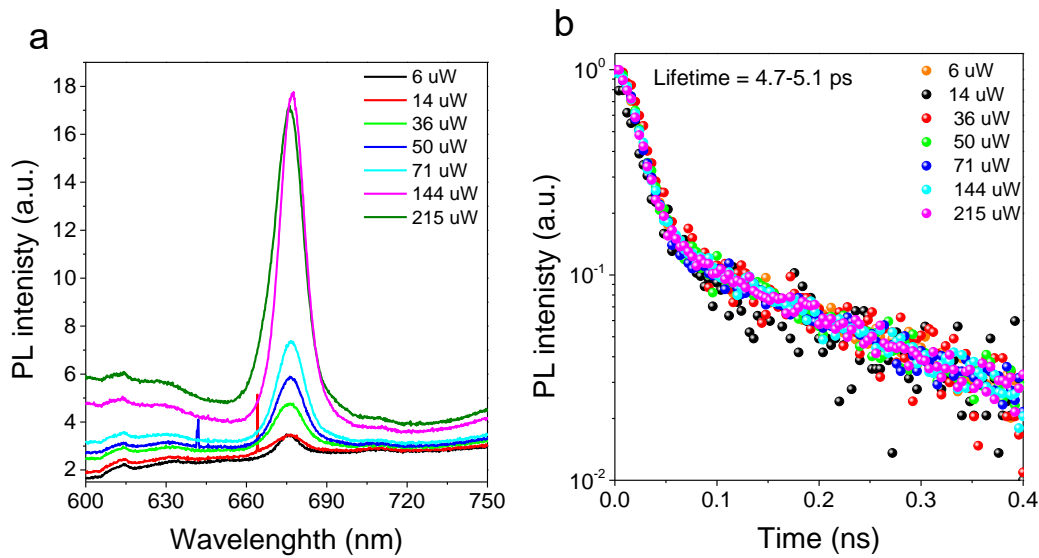

**Figure S11 | Power dependent time-resolved PL at 77 K.** Power dependent measurements at 77 K were carried out on 1L pentacene samples, to confirm the exciton-exciton annihilation effects. We observed an almost constant lifetime (4.7 -5.1 ps) ranging from a very low lase power to a high laser power ranging from 6 μW to 215 μW. The consistent lifetime suggest that the EEA does not play a key role in defining the exciton diffusion characteristics in our sample as shown in the figure below.

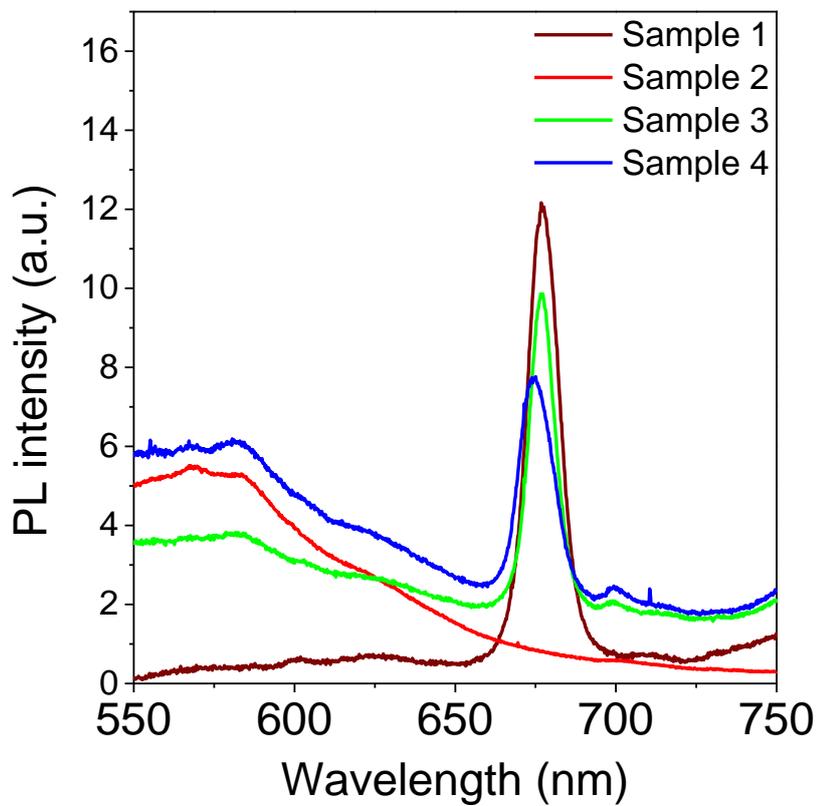

**Figure S12 | Measured PL spectra from various quality of 1L pentacene samples at 77 K,** fabricated under different growth conditions by CVD method.

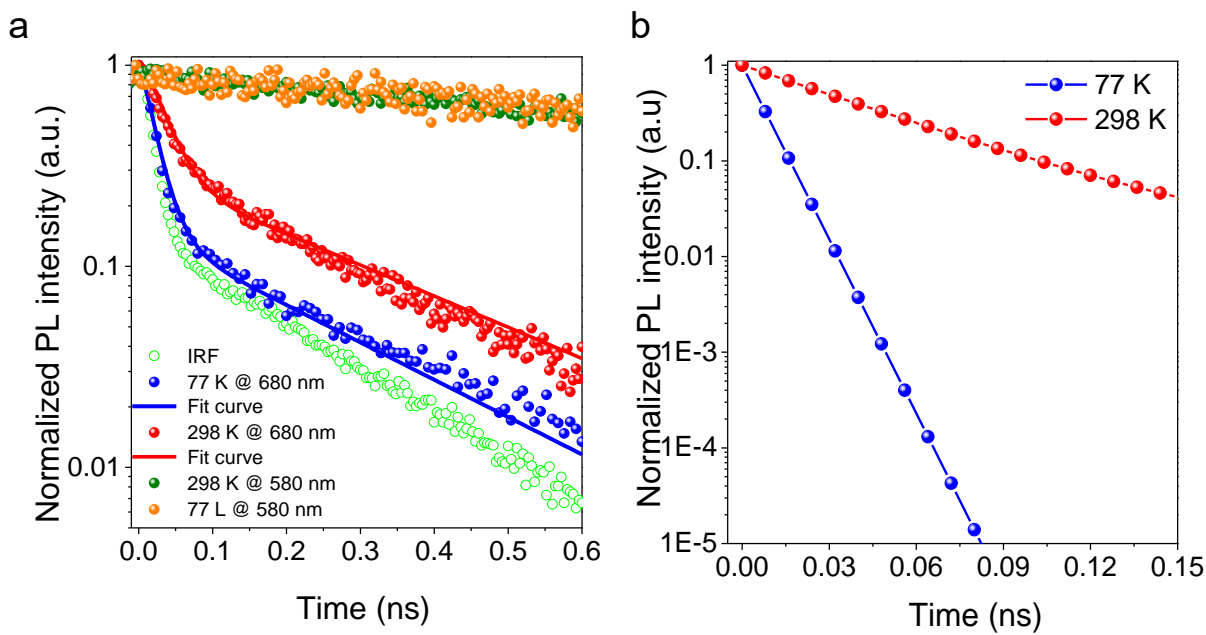

**Figure S13 | Temperature dependent time-resolved PL for low-quality 1L pentacene (sample 3). a**, Time-resolved PL emission (normalized) from 1L pentacene (680 nm, superradiant peak) sample 3 (See Figure S4) at 77 K (blue balls) and 298 K (orange balls). The solid lines represent the fitting curves for both the temperatures. An effective long lifetime of 51.2 ps was extracted from the orange decay curve (298 K) by a fitting with deconvolution using the instrument response function (IRF) (green dots). The lifetime extracted from blue curve (77 K) was 7.6 ps. The drastic reduction in lifetime with decreasing temperature, further supports the superradiance emission from 1L. The effective lifetime at 77 K is slightly higher than what we have reported in Figure 4 (4.1 ps), which may be due to the low crystallinity of the sample 3 as compared to sample 1. **b**, Time-resolved PL emission from the CT excitonic emission at ~580 nm from 1L pentacene sample. The lifetime extracted from the CT peaks is ~2.1 ns and 2.3 ns at 77 K and 298 K respectively. The lifetime is significantly higher as compared to the coherent FR emission peak at 680 nm, due to non-coherent emission from CT excitonic states, which do not show superradiant emission, further supporting the coherent emission visible at 680 nm even from low crystalline samples. **c**, Deconvoluted lifetime curves showing a drastic reduction in lifetime with decreasing temperature.

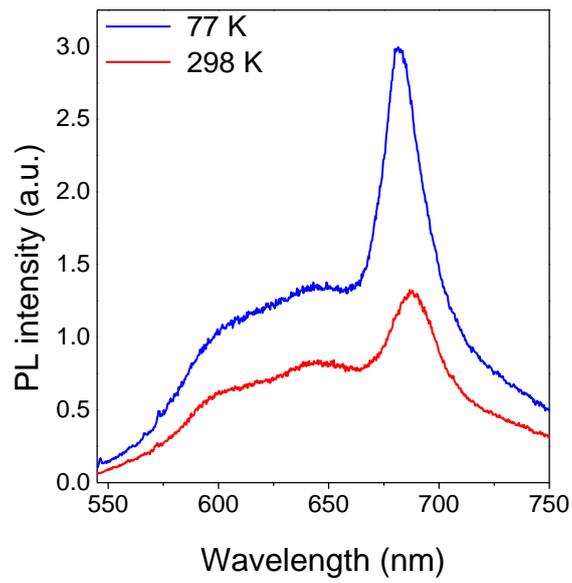

**Figure S14 | Temperature dependent PL measurements from 2L PEN.** PL spectra from 2L regime of PEN samples. The sharp PL emission at 680 nm is visible from 2L but coupled with high-energy CT emissions. The FWHM at 298 K from 2L is ~49 nm and at 77 K is ~30 nm, which is much broader as compared to 1L.

**Table S1: Diffusion coefficients from various organic/in-organic materials: experimental values**

| Material Type | Material Name | Diffusion Coefficient ($cm^2/sec$) | Temperature ($K$) |
|---|---|---|---|
| Inorganic systems | GaAs Quantum wells[42] | 0.1-10 | 4.2 |
| | 1L $MoS_2$[43] | 0.05-0.1 | 298 |
| | 1L $WSe_2$[44] | 14.5 | 298 |
| | 1L $WS_2$[45] | 0.41 | 298 |
| | 1L $MoTe_2$[46] | 0.1 | 70 |
| | Phosphorene [47] | 5 E-3 | 4 |
| | Carbon Nanotubes[48] | 44 | 298 |
| | $WS_2$-tetracene heterostructure[49] | 1 | 298 |
| | $WSe_2$-thin film/bulk[50] | 15/9 | 298 |
| Organic systems | C8S3-J aggregates[27] | 70 | 298 |
| | Light harvesting nanotubes (LHNs-amphiphilic cyanine dyes)[26] | 55 | 298 |
| | Meso-tetra(4-sulfonatophenyl) porphyrin[25] | 3-6 | 298 |
| | Tetracene thin film[51] | 2.8 | 298 |
| | Poly-3 hexylthiophene[52] | 1.8 E-3 | 298 |
| | $C_{70}$ thin film[53] | 3.5 E-3 | 77 |
| | Pentacene thin films[54] | 2.5 E-2 | 298 |
| This work | 2D Pentacene-WL | 2.4 | 298 |
| | | 3.5 | 77 |
| | 2D Pentacene-1L | 346.9 | 298 |
| | | 354.5 | 77 |